\documentstyle[11pt,aaspp4]{article}

\def\om{\Omega }

\def\be{\begin{equation}}
\def\ee{\end{equation}}
\def\eqref#1{(\ref{eqn:#1})}
\def\dg{\delta_g}
\def\dm{\delta_m}
\def\overden{\rho/{\overline \rho}}
\def\overdeng{\rho_g/{\overline \rho}_g}
\def\overdenm{\rho_m/{\overline \rho}_m}
\def\rhobar{{\overline \rho}}
\def\hmpc{{h^{-1}\;{\rm Mpc}}}
\def\hvol{{h^3\;{\rm Mpc}^{-3}}}
\def\hkpc{{h^{-1}\;{\rm kpc}}}
\def\kms{{\rm \;km\;s^{-1}}}
\def\hubunits{\kms\;{\rm Mpc}^{-1}}
\def\kunits{{h\;{\rm Mpc}^{-1}}}

\def\fsp{F_{\rm Sp}}
\def\fs0{F_{\rm S0}}
\def\fe{F_{\rm E}}
\def\divv{\nabla\cdot{\bf v}}

\begin{document}

\title{LOCALLY BIASED GALAXY FORMATION AND LARGE SCALE STRUCTURE}
\author{
\bf Vijay K. Narayanan,
Andreas A. Berlind, and
David H. Weinberg
}
\affil{Department of Astronomy, The Ohio State University, Columbus, OH 43210;}
\affil{ Email: vijay,aberlind,dhw@astronomy.ohio-state.edu}

\begin{abstract}
We examine the influence of the morphology-density relation and a wide range of
simple models for biased galaxy formation on statistical measures of large scale
structure.  We contrast the behavior of local biasing models, in which the 
efficiency of galaxy formation is determined by the density, geometry, or 
velocity dispersion of the local mass distribution, with that 
of non-local biasing models, in which galaxy formation is modulated 
coherently over scales larger than the galaxy correlation length.  If 
morphological segregation of galaxies is governed by a local morphology-density
relation, then the correlation function of E/S0 galaxies should be steeper and
stronger than that of spiral galaxies on small scales, as observed, while on
large scales the E/S0 and spiral galaxies should have correlation 
functions with the same shape but different amplitudes.  Similarly, 
all of our local bias models produce
scale-independent amplification of the correlation function and power spectrum
in the linear and mildly non-linear regimes; only a non-local 
biasing mechanism can alter the shape of the power spectrum on large scales.
Moments of the biased galaxy distribution retain the hierarchical pattern
of the mass moments, but biasing alters the values and scale-dependence of the
hierarchical amplitudes $S_3$ and $S_4$.  Pair-weighted moments of the galaxy
velocity distribution are sensitive to the details of the bias prescription
even if galaxies have the same local velocity distribution as the underlying 
dark matter.  The non-linearity of the relation between galaxy density and
mass density depends on the biasing prescription and the smoothing scale, and
the scatter in this relation is a useful diagnostic of the physical parameters
that determine the bias.  While the assumption that galaxy formation is governed
by local physics leads to some important 
simplifications on large scales, even local biasing is a 
multi-faceted phenomenon whose impact cannot be
described by a single parameter or function.  The sensitivity of galaxy
clustering statistics to the details of galaxy biasing is an obstacle to testing
cosmological models, but it is an equally significant asset for testing 
physical theories of galaxy formation against data from redshift and peculiar 
velocity surveys.
\end{abstract}

\section{Introduction}

Within a decade of defining his classification system for galaxies,
Hubble (1936) observed that elliptical and spiral galaxies tend
to reside in different environments, with elliptical galaxies preferentially
represented in rich clusters and spiral galaxies more numerous in
the field.  More recent studies have confirmed and detailed
this dependence of clustering on morphological type, and they
have shown that luminous galaxies cluster more strongly than faint 
galaxies (e.g., \cite{hamilton88}; \cite{park94}; \cite{loveday95};
\cite{benoist96}; \cite{willmer98}) and that optically selected 
galaxies cluster more strongly than infrared selected galaxies
(e.g., \cite{lahav90}; \cite{saunders92}; \cite{pd94}).
At most one of these galaxy populations can trace the underlying
distribution of mass, and it is more likely that none of them does.
The idea of biased galaxy formation --- preferential formation of
galaxies in high density environments --- was originally introduced
with the hope of reconciling the low apparent mass-to-light ratios of
bound galaxy structures with the theoretically motivated
assumption of an $\Omega=1$ universe (\cite{davis85}; \cite{bardeen86}).
However, it is now broadly recognized that bias, or anti-bias,
is a possibility that we are stuck with, whatever the value of $\om$.
Cosmological N-body simulations can predict statistical properties
of the mass distribution in a specified cosmological model,
but they cannot predict the clustering of galaxies without
an additional prescription for the relation between galaxies and mass.
In this paper, we apply a variety of simple
bias prescriptions to N-body simulations
in order to see how assumptions about bias influence measurable properties of
large scale structure, such as the correlation function, the power
spectrum, moments of galaxy counts, pairwise velocities,
and the relation between galaxy density and peculiar velocity fields.

While the full theory of galaxy formation is likely to be complicated,
a plausible and still very general assumption is that the efficiency
of galaxy formation is determined by properties of the local environment.
For practical purposes, ``local'' means a scale 
comparable to the galaxy correlation length, the distance over which 
material in non-linear structures has mixed during the history of 
the universe.  In a local theory of galaxy formation, when a collapsing
region of the mass distribution decides whether to become a galaxy
(or what type of galaxy to become, or how many galaxies to become),
it has no direct knowledge of material that it has never encountered.
On the other hand, ``non-local'' biasing, in which the efficiency of galaxy
formation varies coherently over larger scales, requires a more
exotic physical mechanism, such as suppression or stimulation
of galaxy formation by ionizing radiation or by some other influence that 
can propagate at the speed of light.
In this paper, we examine several examples of local biasing,
with the efficiency of galaxy formation determined by the
density, pressure, or geometry of the mass distribution in a
surrounding $4\hmpc$ sphere (where $h\equiv H_0/100\hubunits$).
We compare the effects of these local bias models to those of non-local
models inspired by the proposals of Babul \& White (1991)
and Bower et al. (1993).

For purposes of our study, it is important to maintain the 
distinction between a model of biased galaxy formation, which is
a fully specified prescription for populating a mass distribution
with galaxies, and a description of the {\it effects} of bias.
The widely used ``linear bias model'' --- $\dg = b\dm$, where
$\dm \equiv \rho/{\overline\rho}-1$ is the mass density contrast,
$\dg$ is the galaxy density contrast, and $b$ is the bias factor ---
is more properly regarded as a description rather than a model.
One cannot use the $\dg=b\dm$ formula to assign galaxies to
the mass distribution on any scale where the minimum density contrast
is $\delta_{m,{\rm min}} < -1/b$, since it would demand an
unphysically negative galaxy density.  However, linear bias could
emerge as a reasonable approximation to the effects of a physical
bias model on large scales, at least for some purposes.
The relation between the galaxy and mass correlation functions
is often written $\xi_g(r)=b^2(r)\xi_m(r)$, which is
again a description rather than a model.  While it might seem
reasonable to let $b(r)$ be an arbitrary function of scale,
a series of analytic arguments of increasing generality
suggest that for any local bias the quantity $b(r)$ must
asymptote to a constant value in the limit $\xi(r) \ll 1$, making
it impossible for local bias to resolve a discrepancy between the
predicted and observed shapes of $\xi(r)$ on large scales
(\cite{coles93}; \cite{fry93}; \cite{gaztanaga98}; \cite{scherrer98}).
Even the most general of these arguments relies on assumptions about
the mass clustering and the bias model, and it yields only an
asymptotic result.  The numerical approach here complements the analytic
studies by exhibiting fully non-linear solutions for a diverse 
set of biasing models, including some that do not satisfy
the formal assumptions of the analytic calculations.

In principle, the relation between galaxies and mass should be
a prediction of a cosmological theory, not an input to it.
There has been substantial progress towards making such {\it a priori}
predictions
in recent years using hydrodynamic simulations (\cite{co92}, 1993, 1998;
\cite{khw92}, 1996, 1998; \cite{blanton98}),
N-body simulations that attempt to resolve galaxy mass halos
within larger virialized systems (\cite{colin98}),
or a combination of lower resolution N-body simulations
with semi-analytic models of galaxy formation (Kauffmann et al. 1997, 1998;
\cite{governato98}).  However, the semi-analytic models require a 
series of approximations and assumptions, and the numerical studies
suffer from limited resolution and simulation volumes and also
rely on simplified descriptions of star formation and supernova feedback.
At present these approaches yield physically motivated
models of biased galaxy formation that can be tested against
observations, but they do not provide definitive predictions.

In this paper we take a different and in some ways less ambitious tack.
We combine large volume, relatively low resolution N-body simulations
with simple bias prescriptions that are deliberately designed
to ``parametrize ignorance'' about galaxy formation, by illustrating
a wide range of possibilities.  We extend Weinberg's (1995)
study along similar lines by using better simulations and examining
a much wider range of galaxy clustering measures.
Our study also overlaps that of Mann, Peacock, \& Heavens 
(1998, hereafter MPH98),  
who applied several different biasing prescriptions to large N-body
simulations and investigated their influence on the galaxy power spectrum
and on cluster mass-to-light ratios.
Our goal here is to see which measurable properties of galaxy clustering
are sensitive to assumptions about bias and which properties are robust,
or at least are affected in a simple way by local bias.  Our
numerical study complements the general analytic treatment of non-linear,
stochastic biasing models by Dekel \& Lahav (1998, hereafter DL98).

The biasing prescriptions that we adopt in this paper are all
Eulerian models, meaning that the efficiency of galaxy formation
is determined by properties of the evolved mass density field.
These can be contrasted with Lagrangian bias models, such
as the high peaks model (\cite{davis85}; \cite{bardeen86}),
in which the efficiency of galaxy formation depends on the 
initial (linear) density field.  Because galaxies can merge as
structure evolves, neither a pure Lagrangian nor a pure Eulerian
description of bias is entirely realistic.  However, for studies
of galaxy clustering at a specified redshift, either approach may
provide a reasonable approximate description, and the distinction
between them is more technical than physical.
The distinction between local and non-local models, on the other hand,
is fundamentally tied to the physical mechanisms that cause bias.
Eulerian, Lagrangian, and intermediate approaches would yield different
predictions for the evolution of bias with redshift, but we do
not examine this issue in this paper.  The evolution of bias has
been studied using the numerical and semi-analytic approaches
mentioned above and in a more general analytic framework by
Fry (1996) and Tegmark \& Peebles (1998).

We begin our investigation in \S 2 with a simple but informative 
example, in which the galaxy population as a whole traces the mass
but the mix of morphological types is determined by a local
morphology-density relation.  This example illustrates some general
properties of local bias models, and our calculations provide generic
predictions, testable with the next generation of redshift surveys,
for the dependence of the large scale correlation function and
power spectrum on morphological type.  We then move to models in
which the galaxy population has an overall bias, which is applied
using a variety of local and non-local prescriptions.  
We describe the models and examine their influence on a number
of different measures of large scale structure in \S 3.
Section 4 provides a fairly complete summary of our results
and discusses their implications in light of other recent studies
and anticipated observational developments.

\section{The Morphology-Density Relation and the Large Scale Clustering
of Galaxy Types}

The longest recognized and most clearly established form of ``bias''
in the galaxy distribution is the marked preference of early-type
galaxies for dense environments (\cite{hubble36}; \cite{zwicky37}; 
\cite{abell58}).
There have been numerous efforts to quantify
this connection between morphology and environment
(e.g., \cite{dressler80}; \cite{postman84};
\cite{lahav92}; \cite{whitmore93}),
and several groups have used angular correlation functions,
redshift-space correlation functions, or de-projected real-space
correlation functions to characterize the dependence of
clustering on morphology (\cite{davis76}; \cite{giovanelli86}; \cite{loveday95};
\cite{guzzo97}; \cite{willmer98}).
The stronger clustering of early-type galaxies explains the stronger
clustering found in optically selected galaxy catalogs relative
to IRAS-selected catalogs, which preferentially include dusty,
late-type galaxies (\cite{babul90}).
Postman \& Geller (1984, hereafter PG84) parametrize the 
morphology-environment connection as a relation between morphological
fractions and local galaxy density: for overdensities
$\overdeng \la 100$ the fractions of ellipticals, S0s, and
spirals are independent of environment, but at higher densities
the elliptical and S0 fractions climb steadily, eventually saturating
for $\overdeng \ga 10^5$.  
In this Section, we consider a model in
which the galaxy population as a whole traces the underlying mass
distribution but morphological types are assigned according to PG84's
prescription.  Our results will show what should be expected from large
redshift surveys like the 2 Degree Field (2dF) survey (\cite{colless98}) and
the Sloan Digital Sky Survey (SDSS; \cite{gw95}) if a local morphology-density
relation is a correct description of the influence of environment
on morphological type.  This investigation also provides a ``warmup''
for the investigations in \S 3, since the morphology-density relation
is a simple example of a local bias mechanism.

For our underlying mass distribution, we use the output of an N-body
simulation of an open cold dark matter (CDM) model performed by
Cole et al.\ (1998, hereafter CHWF98).  The cosmological parameters are
$\Omega_{m} = 0.4$, $\Omega_{\Lambda}=0$, $\Gamma = 0.25$, $n=1$,
where $\Gamma \equiv \Omega_{m}h \exp(-\Omega_{b} - \Omega_{b}/\Omega_{m})$ 
(\cite{sugiyama95}),
$\Omega_{m}, \Omega_{b}$, $\Omega_{\Lambda}$ 
are the matter, baryon, and vacuum energy density parameters,
and $n=1$ implies a scale-invariant inflationary power spectrum.  
The model is cluster-normalized (\cite{white93}; \cite{eke96}) and
has an rms mass fluctuation $\sigma_8=0.95$ in spheres of 
radius $8\hmpc$.  The amplitude $\sigma_8$ is related to the linear
theory power spectrum $P(k)$ by
\be
 {\sigma}^{2}_{8} = \int_{0}^{\infty} 4\pi k^{2}P(k)\widetilde W^{2}(kR)dk,
\label{eqn:s8def}
\ee
where $\widetilde W(kR)$ is the Fourier transform of a spherical top hat filter
with radius $R=8\hmpc$.
The CHWF98 simulation uses a modified version of the AP3M code 
of Couchman (1991) to follow 
the gravitational evolution of $192^{3}$ particles 
in a periodic cubical box of side $345.6 \hmpc$.
The gravitational force softening is $\epsilon=90\hkpc$ comoving
(for a Plummer force law) and the particle mass is 
$m_p=6.47 \times 10^{11}h^{-1}M_{\odot}$.
Further details of the simulation are given in CHWF98; this is
their model O4S.

Our implementation of the morphology-density relation is based loosely on
figures~1 and~2 of PG84.  We assign each particle in the simulation a
density equal to the mean density in a sphere enclosing its five
nearest neighbors.  We assign the particle a spiral (Sp), S0, or
elliptical (E) morphological type at random, with relative probabilities
$\fsp$, $\fs0$, $\fe$ that depend on the density.  
For $\rho<\rho_F=100\rhobar$, the morphological fractions are 
$\fsp=0.7$, $\fs0=0.2$, $\fe=0.1$.  For 
$\rho_F<\rho<\rho_C=6 \times 10^4\rhobar$,
the fractions are
\begin{eqnarray}
\fsp & = & 0.7-0.6\alpha \nonumber \\
\fs0 & = & 0.2+0.3\alpha \label{eqn:mdrel} \\
\fe  & = & 1-\fsp-\fs0 \nonumber \\
\alpha & = & \log_{10}(\rho/\rho_F)/\log_{10}(\rho/\rho_C). \nonumber
\label{eqn:md}
\end{eqnarray}
For $\rho>\rho_C$, the morphological fractions saturate at
$\fsp=0.1$, $\fs0=0.5$, $\fe=0.4$.
With this formulation,
$13.7 \%$ of the particles are classified 
as ellipticals, $23.7\%$ as S0s, and the remaining $62.6 \%$ as spirals.
Changes to $\rho_F$ or $\rho_C$ or the density assignment method
would alter the numerical values of the bias parameters
discussed below, but they would not change our results in a 
qualitative way.

Figure 1 compares the spatial distributions of the different types of galaxies.
The full galaxy population, a random subset of the N-body particles
with space density $n_g=0.01\hvol$, is shown in panel (a).
For panels (b)-(d), we have artificially boosted the selection
probabilities so that each morphological sub-population has density
$0.01\hvol$; the visual differences among the panels therefore reflect
the differences in the clustering strengths of the 
galaxy types rather than their different number densities.
For our adopted parameters, these visual differences are rather subtle,
though the contrast between clusters and voids is more noticeable
for the ellipticals (panel b) than for the spirals (panel d), as expected.

\begin{figure}
\centerline{
\epsfxsize=\hsize
\epsfbox[18 144 592 718]{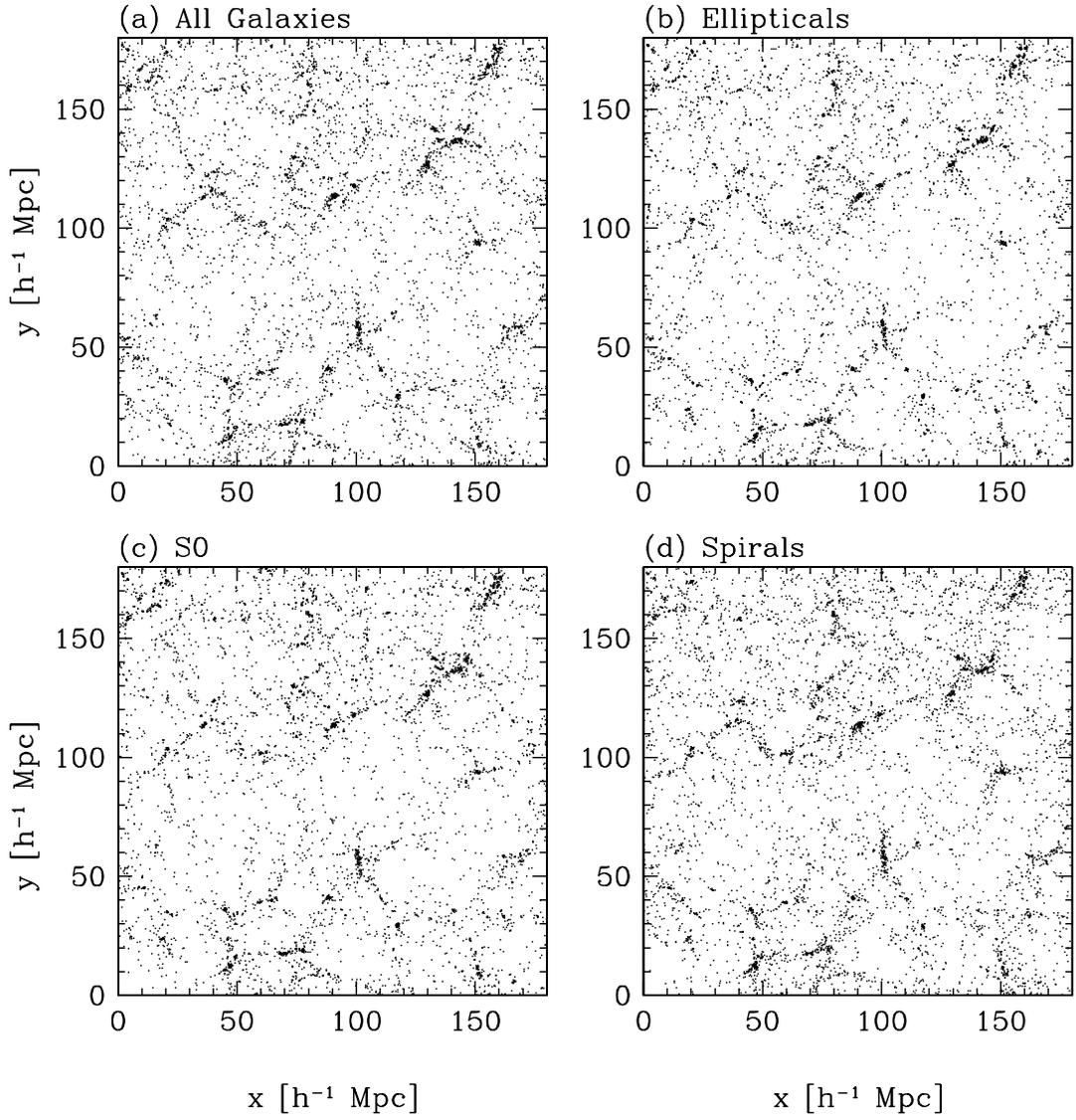}
}
\caption{Galaxy distributions from a local morphology-density relation.
The panels show the galaxy distributions in a slice $18 h^{-1}$Mpc thick
and spanning $180 h^{-1}$Mpc in the other two dimensions.
({\it a}) Underlying mass distribution.
The remaining panels show the galaxy distributions traced by
({\it b}) ellipticals,
({\it c}) S0s, and
({\it d}) spirals.
All of the galaxy distributions have been sampled to the same number density
of $0.01 h^{3}$Mpc$^{-3}$, so that visual differences reflect differences
in clustering rather than differences in galaxy density.
\label{fig:mdgal}
}
\end{figure}

\begin{figure}
\centerline{
\epsfxsize=\hsize
\epsfbox[18 144 592 718]{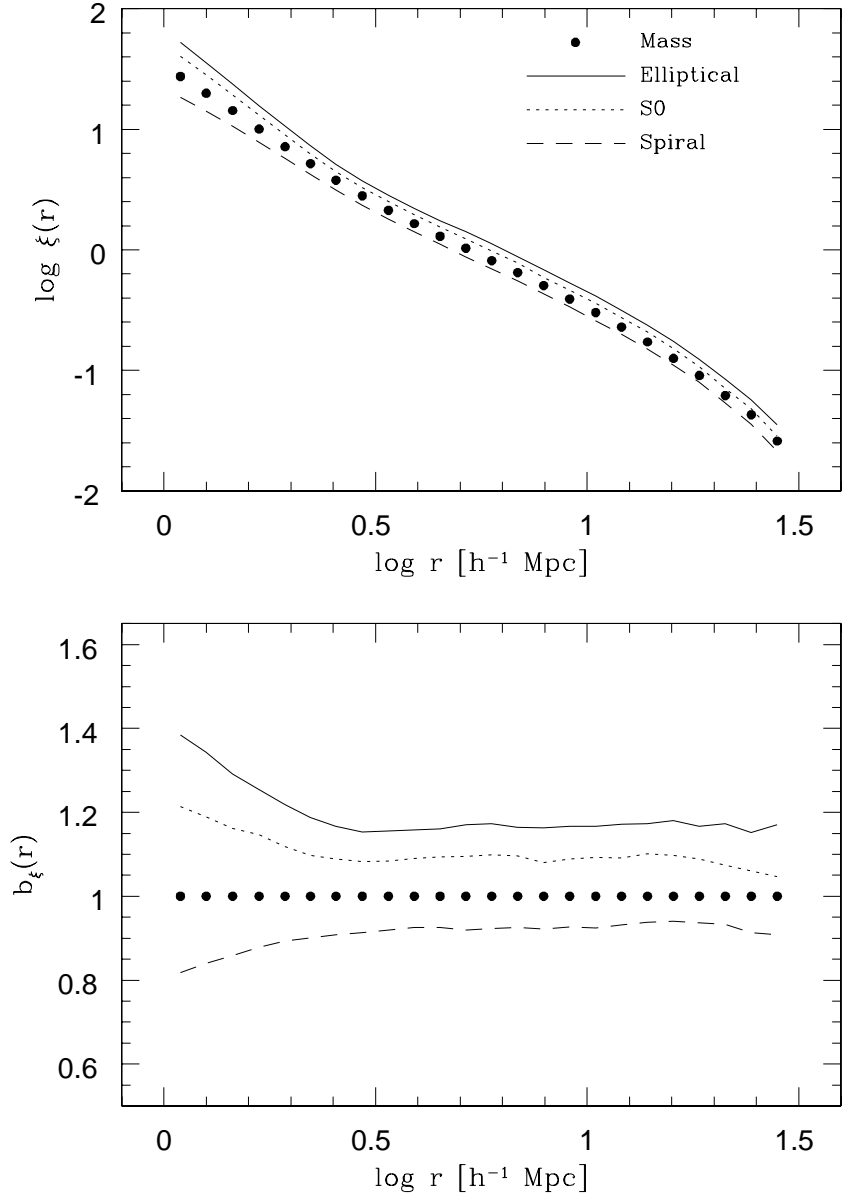}
}
\caption{ Top: Correlation functions of the galaxy 
distributions of different types of galaxies,
with morphological types assigned according to a local
morphology-density relation.
Bottom: The bias functions $b_{\xi}(r)$ defined in equation~(\ref{eqn:brdef}).
\label{fig:mdxi}
}
\end{figure}

Figure~\ref{fig:mdxi} (top panel) compares the two-point correlation
functions $\xi(r)$ of the three galaxy populations to the two-point
correlation function of the mass distribution, which is equal to
that of the full galaxy population by construction.
For $r \la 4\hmpc$, the correlation functions of the E and S0
galaxies are both stronger and steeper than that of the spiral
galaxies, which is just the behavior found in observational
studies of the angular and spatial correlation functions of
different galaxy types (\cite{davis76}; Giovanelli et al.\ 1986; 
\cite{loveday95};
\cite{guzzo97}; Willmer et al.\ 1998).
If one were to fit power laws to these correlation functions and
extrapolate to large $r$, they would cross at $r \sim 10\hmpc$,
with the early-type galaxies more weakly clustered at large scales.
The true behavior in the simulation is quite different, however.
At $r \ga 4\hmpc$, the shape of the $\xi(r)$ of the early-type galaxies 
changes to match that of the spirals, and the amplitude of $\xi(r)$ for 
the early-type galaxies is higher at all scales.  The large scale behavior
is consistent with the observational measurements cited above, but these do not
have sufficient precision at large $r$ to clearly show the E, S0,
and spiral correlation functions tracing each other with a constant
logarithmic offset.  The prediction that the large scale shape of
$\xi(r)$ is independent of morphological type should be easily 
testable with the 2dF and SDSS redshift surveys.

The bottom panel in Figure~\ref{fig:mdxi}
shows the bias function $b_{\xi}(r)$, defined as
\be
b_{\xi}(r) = \sqrt{\frac{\xi_{g}(r)}{\xi_{m}(r)}},
\label{eqn:brdef}
\ee
where $\xi_{g}(r)$  and $\xi_{m}(r)$ are the correlation functions of the 
galaxy and the mass distributions, respectively.
This function becomes independent of scale beyond about $4 \hmpc$, and it 
settles at a different level for each of the galaxy types.
We will show in \S 3 below that this constancy of $b_\xi(r)$ at
large $r$ is a generic feature of local biasing models.
The early-type galaxies are always positively biased with respect
to the mass distribution, $b_{\xi}(r) > 1$, while the spirals 
are anti-biased on all scales, $b_{\xi}(r) < 1$.
The clustering amplitude of the elliptical galaxies on large scales is a 
factor of 1.3 larger than that of the spiral galaxies, 
consistent with the ratio derived using the galaxies in the
Stromlo-APM redshift survey (\cite{loveday95}).
The bias of the early-type galaxies 
increases towards smaller scales, while that of the spirals decreases.
The increase in the relative clustering strength of ellipticals
and spirals towards smaller scales is observed in the clustering 
analysis of the Southern Sky Redshift Survey (Willmer et al.\ 1998).
The bias of the full galaxy population is $b_\xi(r)=1$ by construction.

\begin{figure}
\centerline{
\epsfxsize=\hsize
\epsfbox[18 144 592 718]{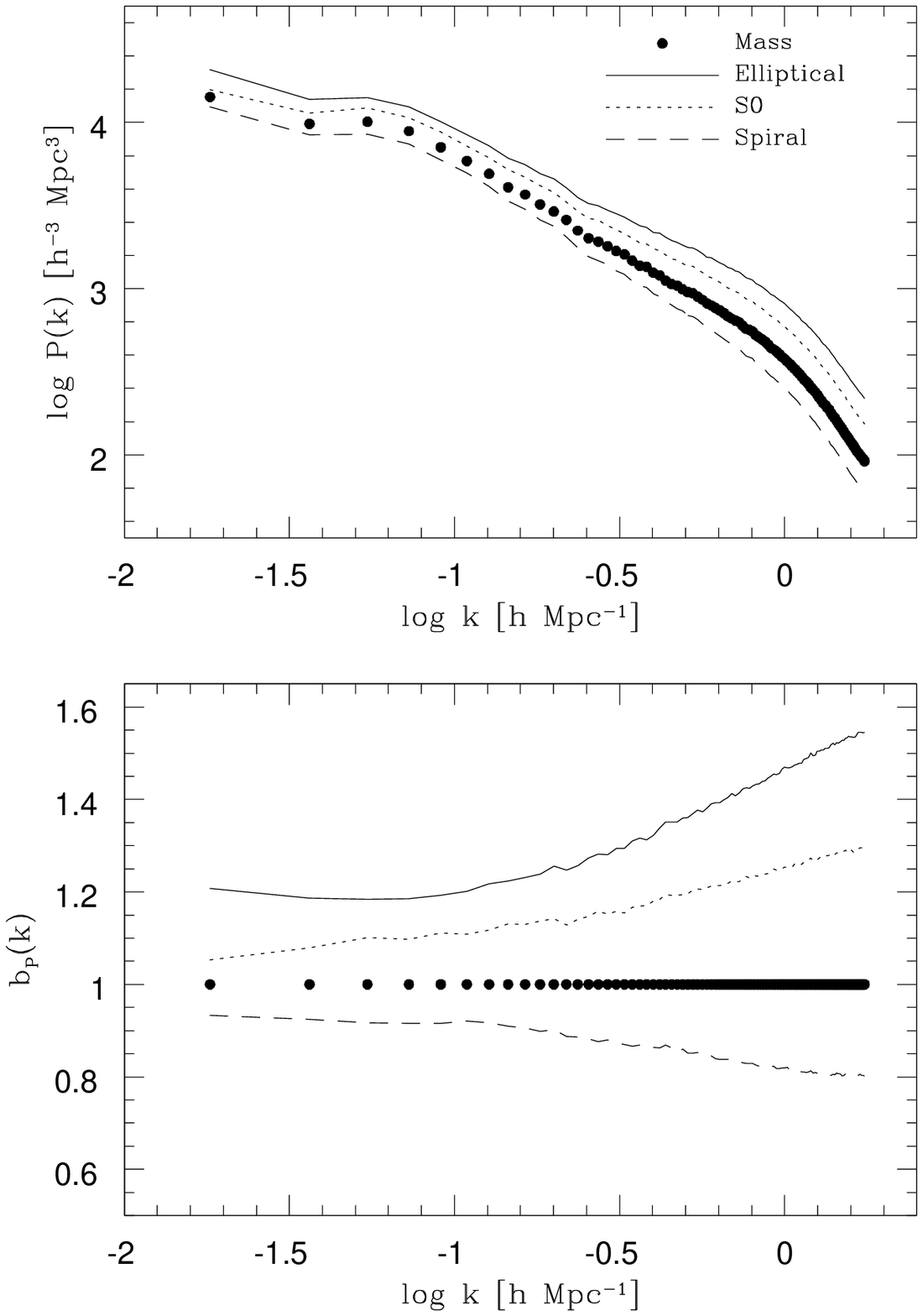}
}
\caption{ Top: Power spectra of the galaxy 
distributions of different types of galaxies,
with morphological types assigned according to a local
morphology-density relation.
Bottom: The bias functions $b_P(k)$ defined in equation~(\ref{eqn:bkdef}).
The largest scale plotted, $k=0.018\kunits$, is the fundamental
mode of the $345.6\hmpc$ simulation cube.
\label{fig:mdpk}
}
\end{figure}

The top panel of Figure~\ref{fig:mdpk} shows the power spectra, $P(k)$, 
of the mass and the galaxy distributions.
We form continuous density fields by cloud-in-cell (CIC) binning 
the discrete particle distributions onto a $192^{3}$ grid,
and compute $P(k)$ using a Fast Fourier Transform (FFT).
We sample all the galaxy distributions to the same number density of 
$0.01 h^{3}$Mpc$^{-3}$, so all the power spectra have the same
shot noise contribution.
The amplification of the clustering of early-type galaxies and 
suppression of the clustering of spiral galaxies is similar to that seen
in Figure~\ref{fig:mdxi}, but the $P(k)$ plot better reveals the behavior
on the largest scales.
The bottom panel shows the bias functions in Fourier space, $b_{P}(k)$, 
defined as
\be
b_{P}(k) = \sqrt{\frac{P_{g}(k)}{P_{m}(k)}},
\label{eqn:bkdef}
\ee
where $P_{g}(k)$  and $P_{m}(k)$ are the power spectra of the galaxy and 
the mass distributions, respectively.
This bias function also becomes independent of scale on large scales and 
has the same relative behavior as $b_{\xi}(r)$ for the different galaxy types.

The assignment of morphological types based on local density produces
a difference in the clustering strength of different galaxy types
at {\it all} scales.  The reason for the large scale bias is the
same one identified by Kaiser (1984) in his discussion of cluster
correlations: regions that are highly overdense on small scales tend
to reside in regions where the large scale, background density
is also high.  The bias functions in Figures~\ref{fig:mdxi}
and~\ref{fig:mdpk} are scale-dependent in the regime where clustering
is non-linear, but they asymptote to constant values on large scales.
We will soon see that this behavior is a generic property of local
biasing models.

We have assumed in this Section that the galaxy population as a whole
traces the mass, but we would expect the relative clustering of
different galaxy types to be similar even if the galaxy population
has a net bias or anti-bias, at least if the mechanism that produces
that bias is local.  While the numerical values of the relative
bias parameters depend on our choices of $\rho_F$ and $\rho_C$
(eq.~[\ref{eqn:mdrel}]) and our method of defining the local density,
the qualitative prediction that the correlation functions and
power spectra of E, S0, and spiral galaxies have the same
shape at large scales should hold in any model where the morphological
type is determined by the local density.  

\section{The Influence of Bias on Galaxy Clustering}

If the environment influences the formation efficiency of individual
galaxy types, then it is likely to influence the overall efficiency
of galaxy formation as well.  While we are far from having a
definitive theoretical account of galaxy formation, the numerical
and semi-analytic studies cited in \S 1 suggest that galaxies should 
be significantly biased tracers of structure, at least in some regimes
of luminosity and redshift.
With the morphology-density results as background, we now turn to
a much broader set of biasing models, some local and some non-local.
We no longer attempt to discriminate among galaxy types but instead
consider models in which the galaxy population as a whole is biased
with respect to the underlying mass distribution.

For our underlying model (described in \S 3.1), we adopt an
$\Omega=1$, tilted CDM model.  This model has an rms mass fluctuation
amplitude $\sigma_8=0.55$, about a factor of two below the observed
fluctuation amplitude of bright optical galaxies (e.g., \cite{dp83};
\cite{loveday95}).  
We apply a number of different local and non-local
Eulerian biasing prescriptions (described in \S 3.2 and \S 3.3)
to this model, each of which has a single adjustable parameter that
controls the strength of the bias.  We choose the value of this
parameter by requiring that the ratio of rms galaxy fluctuations
to rms mass fluctuations,
\be
b_{\sigma} = \frac{\sigma_{g}(R)}{\sigma_{m}(R)},
\label{eqn:bdef}
\ee
is $b_\sigma=2$ in spheres of radius $R = 12 \hmpc$.
The definition of the bias factor in equation~\eqref{bdef} 
is identical to the quantity $b_{\rm var}$ defined by DL98.
We fix the bias factor at $R=12\hmpc$ rather than $R=8\hmpc$ so
that our local bias models have similar bias at very large scales.
We sample the biased galaxy distributions to an average density of 
$0.01 h^{3}{\rm Mpc}^{-3}$, so that all of the galaxy distributions have 
comparable shot-noise properties.

\subsection{Mass Distribution}

In all of the Eulerian biasing schemes that we study in this paper, we choose 
the galaxies from the underlying mass distribution of a 
tilted CDM model with $\Omega_{m} = 1, \Omega_{\Lambda} = 0$, and 
$\Omega_{b} = 0.05$.
The primordial slope of the power spectrum, $n$, is adjusted to 
simultaneously match the CMB anisotropies on large scales and 
the observed masses of galaxy clusters on small scales.
The cluster constraint requires
$\sigma_{8}=0.55$ (White et al.\ 1993), which implies $n = 0.803$ 
if one incorporates the standard inflationary prediction for
gravitational wave contributions to the COBE anisotropies.
We compute the CDM transfer function using the analytical fitting functions
provided by Eisenstein \& Hu (1998).

Our model parameters are the same as those of the E2 model of 
Cole et al.\ (1997, see also CHWF98).
However, for the purposes of our investigation it is more important
to have good statistics on very large scales than to have high 
gravitational force resolution.  We therefore perform our own N-body
simulations of this model using a 
particle-mesh (PM) N-body code written by C. Park, which is
described and tested by Park (1990; see also \cite{park91}).
Each simulation uses $200^{3}$ particles and a $400^{3}$ force mesh 
to follow the gravitational evolution in a periodic cube $400\hmpc$
on a side.
We start the gravitational evolution from a redshift $z = 20$ and follow
it to $z = 0$ in 80 equal incremental steps of the expansion scale factor 
$a(t)$.
We also evolved this mass distribution using twice this number of time steps
and found that the correlation function and the velocity dispersion of the 
resulting mass distributions changed very little, showing that our results 
are robust to increasing the temporal resolution of the N-body
simulation.
We ran four such simulations with different random phases for the 
Fourier components of the initial density field, and the results
we show are averaged over these four independent mass/galaxy distributions.

\subsection{Local Biasing Schemes}

We now describe our local Eulerian biasing models in detail.
The first two schemes, density-threshold bias and power-law
bias, select galaxies based on the local mass density alone. 
The sheet bias scheme selects galaxies based on the geometry
of the local mass distribution.
We also investigated two other local bias schemes, one based on pressure
and one on geometry, that we will omit from our detailed presentation
of results because they prove very similar to the density-threshold
and sheet bias schemes, respectively.
We compute all the local properties associated with a mass particle,
including the local mass density and the moment of inertia tensor, 
in a sphere of radius $4h^{-1}$Mpc around that particle.

\subsubsection{Density-threshold bias}

In order for galaxies to have a net bias $b_\sigma>1$, they should
preferentially populate regions of higher mass density.
The simplest prescription that achieves this is a density-threshold bias:
galaxy formation is entirely suppressed below some threshold, and galaxies
form with equal efficiency per unit mass in all regions above the
threshold.  This biasing scheme was adopted in some of the early
numerical investigations of CDM models (e.g., \cite{melott86}), and it has
been used extensively by J.\ Einasto and collaborators in theoretical
modeling of voids and superclusters (e.g., \cite{einasto94}).
In the density-threshold bias model, the probability that a particle
with local mass density $\rho_{m}$ is selected as a galaxy is 
\be
P = \cases{ A &if $\rho_{m} \ge B,$ \cr
            0 &if  $\rho_{m} < B$. \cr }
\label{eqn:denbias}
\ee
We choose the threshold density $B$ so that the bias
factor $b_{\sigma}(12\hmpc)=2$ (eq.~[\ref{eqn:bdef}]) and the
probability $A$ so that the mean galaxy density is $n_g=0.01\hvol$.
        
\subsubsection{Power-law bias}

The threshold model is extreme in the sense that galaxy formation is
completely suppressed at low densities and independent of density
above the threshold.  In our second model, we make the galaxy
density a steadily increasing, power-law function of the mass density,
$(\overdeng) \propto (\overdenm)^B$.  The selection probability for a 
particle with local mass density $\rho_m$ to become a galaxy is therefore
\be
P = A(\overdenm)^{B-1} .
\label{eqn:plbias}
\ee
We choose the value of $B$ so that $b_\sigma=2$ and the value of $A$
so that $n_{g} = 0.01h^{3}$Mpc$^{-3}$.
This biasing relation is similar to the one suggested by Cen \& Ostriker 
(1993) based on hydrodynamic simulations incorporating physical models
for galaxy formation (\cite{co92}), but it differs in that there
is no quadratic term that saturates the biasing relation at 
high mass densities.
Little \& Weinberg (1994) have compared the influence of 
density-threshold bias and Cen-Ostriker bias on the void probability
function, showing that the size of empty voids is substantially
larger for threshold bias at fixed $b_\sigma$.  MPH98 included
Cen-Ostriker bias in their general study of Eulerian bias schemes,
and CHWF98 employ both density-threshold bias and
power-law bias in their mock catalogs of the 2dF and SDSS redshift surveys.

\subsubsection{Sheet bias}

The two biasing schemes described above choose galaxies with a probability
that is some function of the local mass density alone.
In principle, we could envisage other local properties that can influence
the efficiency of galaxy formation.
Redshift surveys of the local universe reveal a striking pattern 
in the galaxy distribution, with a large number of galaxies lying in thin
walls and narrow filaments on the periphery of huge voids 
(\cite{delapparent86}).
The process of gravitational condensation and cooling could be
substantially different in a structure that is effectively 2-dimensional
rather than 3-dimensional (e.g., \cite{ostriker81}; \cite{vishniac85};
\cite{white90}).  These considerations, and the desire to investigate
a radical alternative to density-based models,
motivate us to consider a biasing scheme in which the efficiency
of galaxy formation depends solely on the geometry of the local mass 
distribution.

In order to identify sheet-like regions of the mass distribution,
we compute the three eigenvalues $\lambda_3>\lambda_2>\lambda_1$
of the moment of inertia tensor in the $4\hmpc$ sphere
surrounding each N-body particle.
The selection probability for a particle to be a galaxy is
\be
P = \cases{ A &if $\lambda_3/\lambda_1 \ge B$ \cr
            0 &if  $\lambda_3/\lambda_1 < B$. \cr }
\label{eqn:sheetbias}
\ee
This procedure selects galaxies in regions where the mass distribution
is planar, $\lambda_3 \gg \lambda_1$, and avoids regions where
the mass distribution is isotropic, $\lambda_3 \approx \lambda_1$.
The flatness ratio $\lambda_3/\lambda_1$ is correlated with the
mass density, so raising the threshold $B$ increases the bias 
factor $b_\sigma$.  We choose the value of $B$ so that 
$b_\sigma(12\hmpc)=2$ and the value of $A$ so that $n_g=0.01\hvol$.
Because the density and the flatness ratio are not perfectly correlated, 
the sheet bias scheme differs significantly from the density-threshold bias 
scheme.
In fact, one cannot obtain an arbitrarily high bias simply by raising the
value of $B$ in equation~\eqref{sheetbias}, but for our adopted
parameters we find that we can always achieve $b_\sigma=2$.
We eliminate all particles that have fewer than 18 neighbors
within $4\hmpc$ ($\overdenm < 0.55$) because the moment-of-inertia
tensor would be too noisy.

\subsubsection{Pressure and filament bias}

We also investigated two other local biasing schemes considered
by Weinberg (1995).  The first of these, pressure bias, is similar
to density-threshold bias, except that the galaxies are selected
based on a threshold in $\rho_m\sigma_v^2$ rather than $\rho_m$,
where $\sigma_v^2$ is the velocity dispersion in a $4\hmpc$ sphere.
This scheme is motivated by the possibility that the pressure of
the local intergalactic medium could play a role in stimulating
galaxy formation.  However, the velocity dispersion $\sigma_v^2$
is itself fairly well correlated with $\rho_m$, and we decided to 
omit results for the pressure bias model from our figures below
because they are nearly identical to those of the density-threshold
bias model (as also found by Weinberg 1995).
This similarity of results seems somewhat at odds with 
Blanton et al.'s (1998) finding that the inclusion of gas
temperature or dark matter velocity dispersion as a variable in
the bias relation increases the scale-dependence of the bias factor.
We return to this issue in \S 4.

The second additional scheme, filament bias, is similar to sheet bias
(eq.~[\ref{eqn:sheetbias}]),
except that the selection is based on the eigenvalue ratio 
$\lambda_3/\lambda_2$ instead of $\lambda_3/\lambda_1$.
Our quantitative results for filament bias are somewhat different
from those of sheet bias, but they are similar enough that we decided
not to include them as separate curves or panels in our figures.
In the case of an identical threshold flatness ratio $B$,
the particles selected by filament bias are a subset of those
selected by sheet bias.  However, a given value of $B$ produces different
values of $b_\sigma$ for sheet and filament bias.

\subsection{Non-local Biasing Schemes}

In addition to the local biasing schemes described in \S 3.2, we
will investigate two non-local biasing schemes, inspired by the
models of Babul \& White (1991, hereafter BW91) and Bower et al.\ (1993,
hereafter BCFW93).  These papers were a response to measurements of
angular clustering in the APM galaxy catalog (\cite{maddox90}),
which showed that the ``standard'' CDM model 
(SCDM, with $\Omega=1$, $h=0.5$, $n=1$) predicted the wrong
power spectrum shape on large scales.
BW91 showed that the SCDM power spectrum could be reconciled with
the APM measurements if the formation of luminous galaxies was
suppressed in randomly placed spheres with a filling factor 
$f \sim 0.7$ and radii $R \sim 15\hmpc$, perhaps because of
photoionization by quasars (\cite{dekel87}; \cite{braun88}).
BCFW93 achieved a similar result with a scheme that modulates
galaxy luminosities less drastically but over larger scales
({\it Gaussian} filter radii $R \sim 20\hmpc$).  
Both papers argue that the discrepancy between SCDM and the APM
data could arise from galaxy formation physics rather than a fundamental
failing of the cosmological model.
However, while both papers introduced non-local bias models, 
neither addressed the question of whether non-locality was essential
to achieving the desired modulation of the large scale galaxy 
correlation function.  Subsequent analytic arguments have suggested
that non-locality is indeed essential (\cite{coles93}; \cite{fry93};
\cite{gaztanaga98}; \cite{scherrer98}), and our results below
will strengthen the case.

In our first scheme, which we will refer to simply as ``non-local bias,''
we select galaxies with a probability
\be
P_{\rm nl} = P_{\rm l}(1 + \kappa \bar \nu),
\label{eqn:nlbiasdef}
\ee
where $P_{\rm nl}$ and $P_{\rm l}$ refer respectively to the probability
of selecting a mass particle to be a galaxy in the presence and absence 
of non-local effects.
We model the local probability $P_{\rm l}$ using the power-law
scheme described in \S3.2.2.
We fix the index of this power law, $B$, by requiring that $b_\sigma=2$ 
on a scale of $12 \hmpc$ and the probability $A$ so that 
$n_{g} = 0.01h^{3}$Mpc$^{-3}$.
The non-locality is introduced through the dependence of $P_{\rm nl}$ on
$\bar \nu$, which is the density in a top hat sphere of radius $R_{\rm nl}$
about the particle, in units of the rms mass fluctuation on this scale.
The large scale smoothing radius $R_{\rm nl}$ defines 
the scale of the non-local influence, 
and the modulation coefficient $\kappa$ controls its strength.
Here, we choose $R_{\rm nl} = 30 \hmpc$ and $\kappa = 0.25$  for purely
illustrational purposes, guided mainly by the fact that non-local effects
on this scale can reconcile the SCDM model with the 
APM correlation function (BCFW93; note our use of a top hat filter
rather than a Gaussian filter).
While the non-local biasing scheme resembles BCFW93's 
cooperative galaxy formation model, our implementation is quite different
in detail --- Eulerian instead of Lagrangian, with power-law bias taking
the place of high peak bias.

In our second non-local scheme, which we will refer to as ``void bias,''
we randomly place spheres of radius $R_v=15\hmpc$ and eliminate all particles
lying within these spheres.  We randomly select galaxies from the 
remaining particles so that $n_{g} = 0.01h^{3}$Mpc$^{-3}$.
We choose the filling factor of the voids
so that $b_\sigma=2$ at $R=12\hmpc$.  This model is similar to that of BW91,
who suggested that the voids might be produced by photoionization
of pre-galactic gas by quasars.

Although one could easily construct many other non-local bias prescriptions,
these two models will suffice to illustrate the differing effects of
local and non-local bias.  We regard non-local bias as {\it a priori}
less plausible than local bias because of the difficulty of 
producing coherent modulations in galaxy properties over such large scales.
A wide-ranging exploration of non-local models therefore seems justified
only if future observational developments force us to consider them 
more seriously.

\subsection{Galaxy Distributions}

\begin{figure}
\centerline{
\epsfxsize=\hsize
\epsfbox[18 144 592 718]{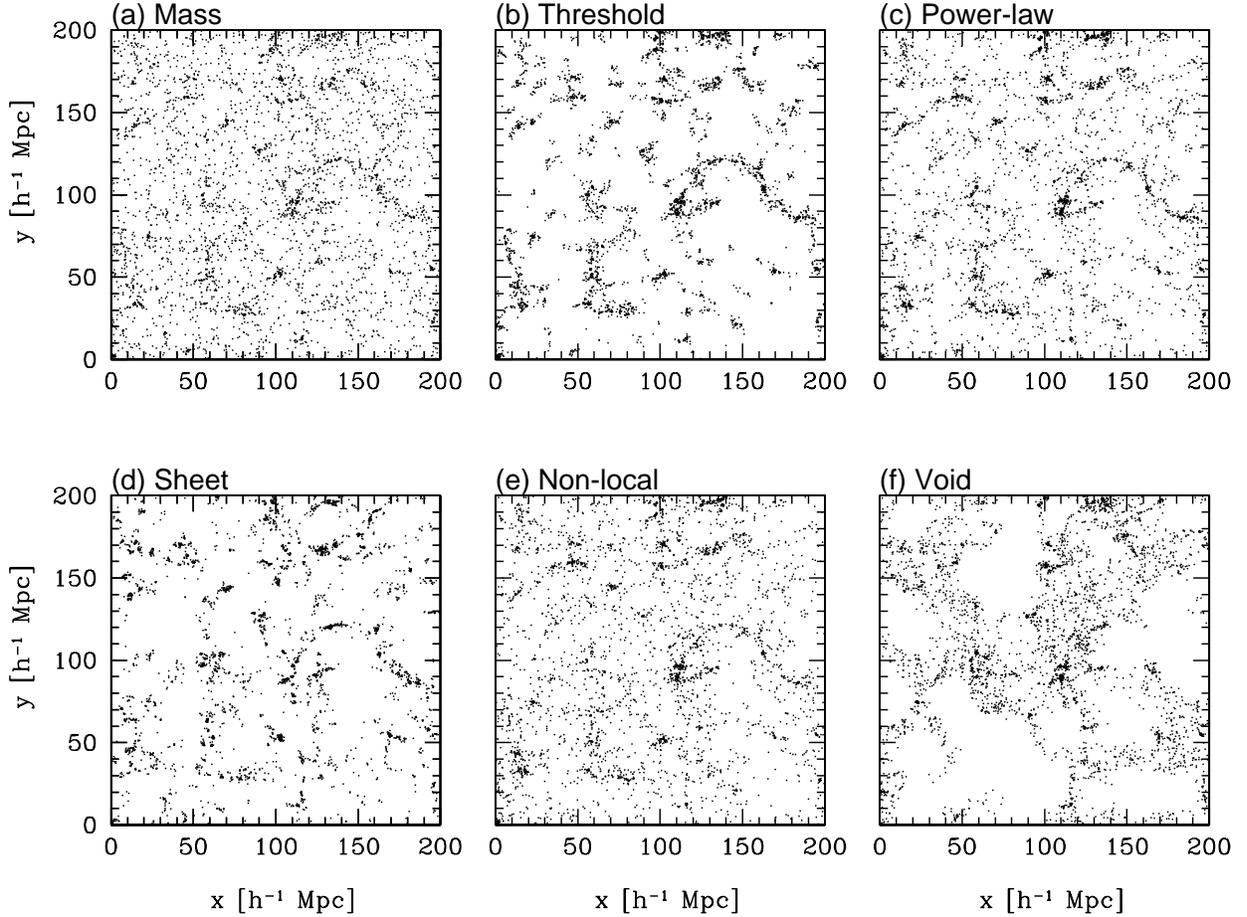}
}
\caption{
Galaxy distributions under the different biasing schemes.
The panels show the galaxy distributions in a slice $20 h^{-1}$Mpc thick
and spanning $200 h^{-1}$Mpc in the other two dimensions.
({\it a}) Underlying mass distribution, randomly sampled to $n_g=0.01\hvol$.
The remaining panels show the galaxy distributions under
({\it b}) density-threshold bias,
({\it c}) power-law bias,
({\it d}) sheet bias,
({\it e}) non-local bias, and
({\it f}) void bias.
\label{fig:galdist}
}
\end{figure}

Figure~\ref{fig:galdist}
shows the mass distribution (in panel a) and the various biased
galaxy distributions from one of our simulations.
We plot the locations of all the galaxy particles that lie in a 
region $20 \hmpc$ thick about the center of the cube.
The galaxy distributions appear strikingly different
from each other even though all of them have the same bias
factor $b_\sigma=2$ at $R=12\hmpc$.
Density-threshold bias (panel b) completely eliminates galaxies in the
low density regions, leaving  only the peaks and filaments of the 
mass distribution populated by galaxies.
The power-law bias distribution
(panel c) looks like a more dynamically evolved
version of the mass distribution, as the contrast between the overdense 
and the underdense regions is enhanced.
However, the underdense regions are not completely devoid of 
galaxies as they are in the threshold bias model.
The sheet bias scheme (panel d) preferentially selects galaxies lying in
anisotropic structures, avoiding both the underdense regions and
the more isotropic clusters that are so prominent in the power-law model.
In this two-dimensional representation, the galaxies appear to lie on 
thin elongated filaments, with very few knot-like features.
The galaxy distribution of the non-local bias model (panel e)
looks remarkably similar to that of the local power-law bias model, 
although they have very 
different large scale clustering properties as we will see below.
The void bias scheme (panel f) completely wipes out all galaxies in some 
regions.  The void radius $R_v$ imprints an obvious 
characteristic length scale on the galaxy distribution.

\subsection{Correlation Function and Power Spectrum}

\begin{figure}
\centerline{
\epsfxsize=\hsize
\epsfbox[18 144 592 718]{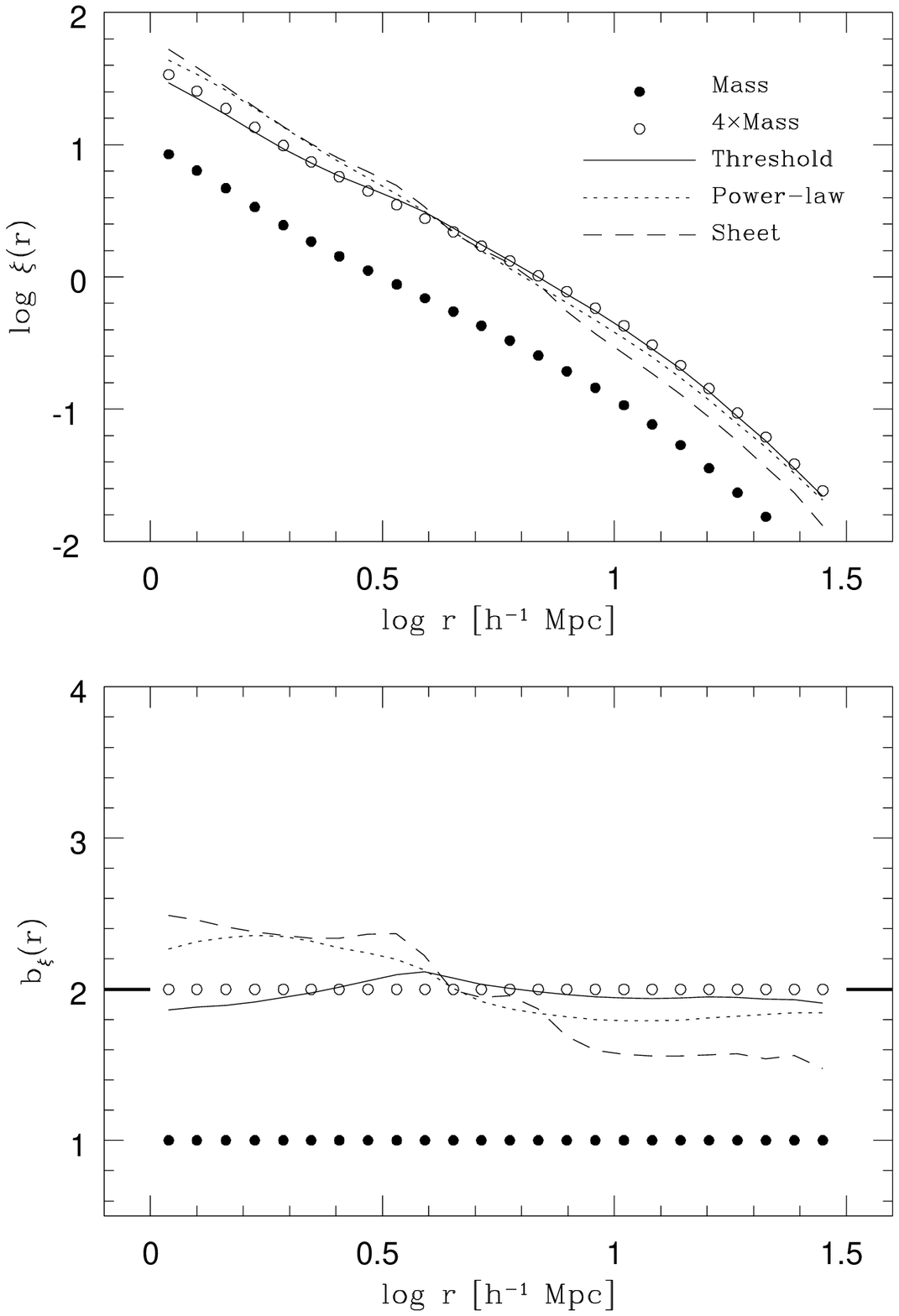}
}
\caption{
Top: Correlation functions of the galaxy distributions under the different 
local biasing schemes: density-threshold bias (solid), power-law bias
(dotted), and sheet bias (dashed).  Filled circles show the mass
correlation function, and open circles show the mass correlation 
function multiplied by $b_\sigma^2(12\hmpc)=4$.
Bottom: The bias functions $b_{\xi}(r)$ defined in equation~(\ref{eqn:brdef}).
\label{fig:xi}
}
\end{figure}

Figure~\ref{fig:xi} shows the correlation functions $\xi(r)$
and the bias functions $b_\xi(r)$ of
the different galaxy distributions under the local biasing schemes.
Our normalization condition $\sigma_g(12\hmpc)=2\sigma_m(12\hmpc)$
imposes an integral constraint on $\xi_g(r)$ that is, approximately,
$\int_0^{12} r^2 \xi(r) dr = {\rm constant}$.  The bias models
that produce higher $\xi(r)$ on small scales must therefore compensate
with lower $\xi(r)$ on large scales.  However, in all three cases,
the correlation bias function $b_\xi(r)$ becomes independent of scale
for $r > 8\hmpc$, and for the two density-based schemes $b_\xi(r)$
is nearly scale-independent for $r>5\hmpc$.
As noted in \S 3.2, results for a pressure-threshold bias are
similar to those for density-threshold bias, and results for
filament bias are similar to those for sheet bias.  We also
found scale-independent large scale amplification of $\xi(r)$
for the morphology-density relation (Figure~\ref{fig:mdxi}), which
is itself a form of local bias.

Our results for the density-based bias schemes are as expected in light
of the analytic argument by Scherrer \& Weinberg (1998), which shows
that $b_\xi(r)$ tends to a constant in the limit $\xi(r) \ll 1$
for any local density bias applied to a field with a hierarchical
clustering pattern.  At least in the examples we have examined,
the condition $\xi(r)<1$ seems to be sufficient to reach the asymptotic 
regime.  More importantly, we find the same asymptotically constant
behavior of $b_\xi(r)$ for sheet, pressure, and filament bias, where
the galaxy density is not a 
simple function of the mass density.  While we cannot examine
all conceivable local bias schemes in a finite set of numerical
experiments, the combination of our results with the general analytic
arguments for local density schemes strongly suggests that
scale-independent bias in the regime $\xi(r)<1$ is a generic property
of all local models of galaxy formation.

\begin{figure}
\centerline{
\epsfxsize=\hsize
\epsfbox[18 144 592 718]{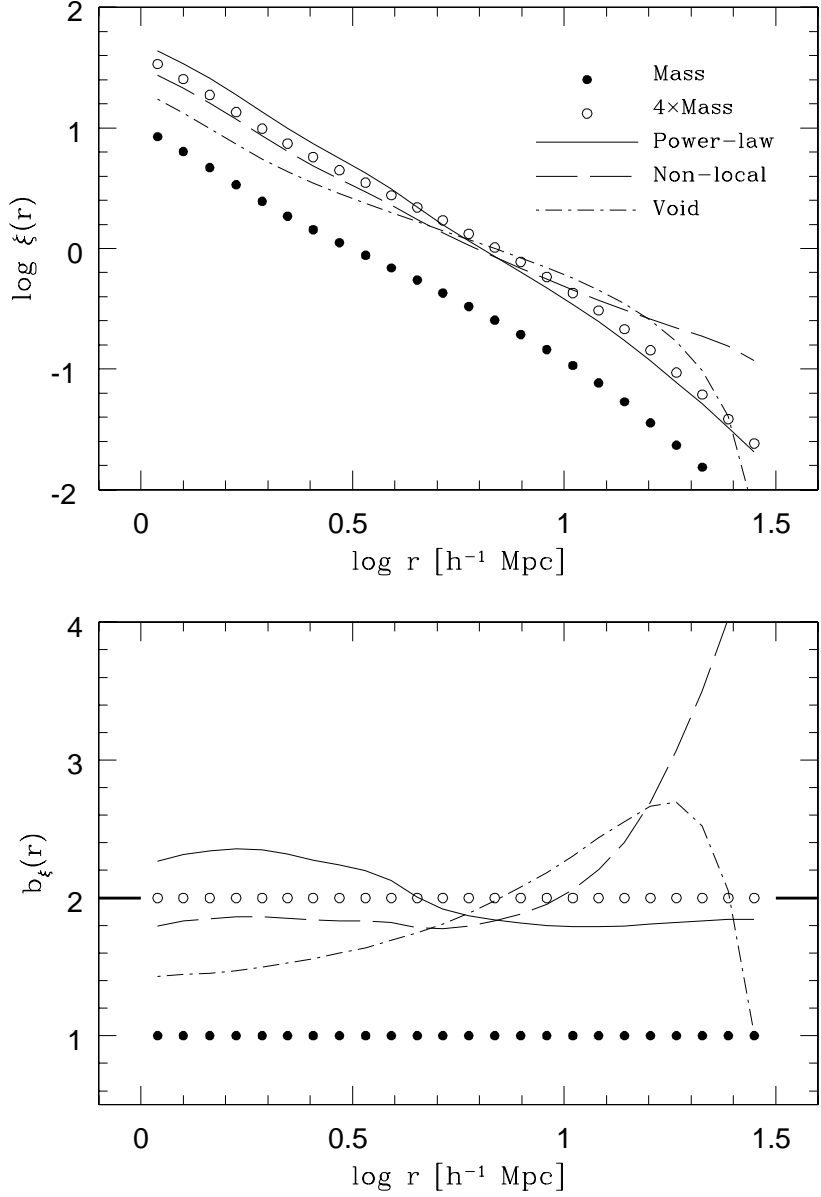}
}
\caption{
Top: Correlation functions of the galaxy distributions under the power-law
(solid), non-local (long-dashed), and void (dot-dashed) biasing schemes.
Bottom: The bias functions $b_{\xi}(r)$ defined in equation~(\ref{eqn:brdef}).
\label{fig:nlxi}
}
\end{figure}

Figure~\ref{fig:nlxi}
shows correlation functions and bias functions in the same format
as Figure 5, but for the two non-local biasing schemes described
in \S 3.3, ``non-local bias'' and ``void bias.''
We also show the results of the local power-law biasing scheme 
(from Figure~\ref{fig:xi}) for comparison.
The non-local bias scheme clearly leads to enhanced clustering on scales larger
than about $10 \hmpc$.
The bias function increases monotonically up to the largest scale
shown in the Figure, growing
by almost a factor of two between $10 \hmpc$ and $30 \hmpc$.
Void bias boosts the correlation function substantially on scales
$r \sim R_v \sim 15\hmpc$, while at larger scales
the imprint of empty regions with a characteristic diameter
causes $\xi(r)$ to turn over rapidly.
Our results confirm the arguments of BW91 and BCFW93 that large scale
modulations of galaxy formation can alter the shape of $\xi(r)$
enough to reconcile a standard CDM mass correlation function with
the APM galaxy correlation function.  However, the results in
Figure~\ref{fig:xi} imply that non-locality is not an incidental
feature of the BW91 and BCFW93 models but is essential to obtaining
a scale-dependent $b_\xi(r)$ at large $r$.  

\begin{figure}
\centerline{
\epsfxsize=\hsize
\epsfbox[18 144 592 718]{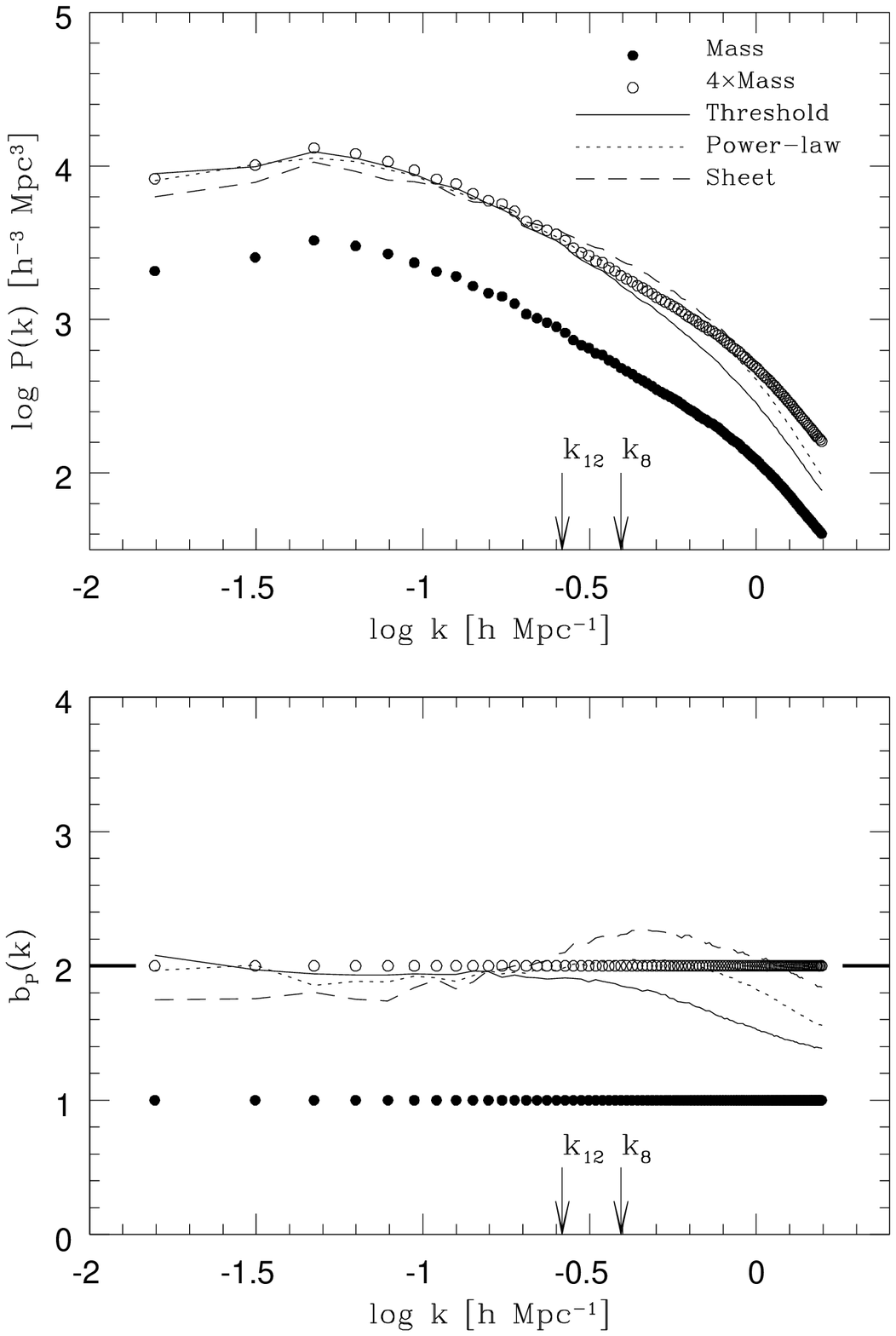}
}
\caption{
Top:  Power spectra of the galaxy distributions under the different local 
biasing schemes.
Bottom:  The bias functions $b_{P}(k)$ defined in equation~(\ref{eqn:bkdef}).
\label{fig:pk}
}
\end{figure}

Figure~\ref{fig:pk}
shows the power spectra $P(k)$ and the Fourier space
bias functions $b_{P}(k)$ for the local biasing schemes.
The arrows marked $k_{8} = 2\pi/(2 \times 8) = 0.3927 h$ Mpc$^{-1}$ and 
$k_{12} = 2\pi/(2 \times 12) = 0.2618 h$ Mpc$^{-1}$ represent the wavenumbers 
corresponding to the length scales $8 \hmpc$ and $12 \hmpc$, at which we 
normalize the amplitude of the mass power spectrum and the bias 
factor, respectively.
This bias function $b_{P}(k)$ is also independent of scale on large scales,
where it is approximately equal to $b_\sigma(12\hmpc)=2$.
This Figure quantifies clustering on scales up to the fundamental
wavelength of our $400\hmpc$ simulation cube, well beyond
those probed by the correlation functions in Figure~\ref{fig:xi}.
The constancy of $b_{P}(k)$ at small $k$ reinforces our
conclusion that the bias functions of local biasing schemes 
remain scale-independent on {\it all} large scales.
The bias functions become truly scale-independent 
only for $k<k_{s} = 0.2 h$ Mpc$^{-1}$ 
[corresponding to length scales $R > 2\pi/(2 \times k_{s}) 
\approx 15 \hmpc$],
although our binning in Fourier space at the low wavenumbers is 
admittedly coarse.
This value of $k_{s}$ is comparable to the fundamental frequency of the 
simulation box used by MPH98,
who found a mild, monotonic scale-dependence of
$b_{P}(k)$ on smaller scales ($k > 0.1 h$ Mpc$^{-1}$).

\begin{figure}
\centerline{
\epsfxsize=\hsize
\epsfbox[18 144 592 718]{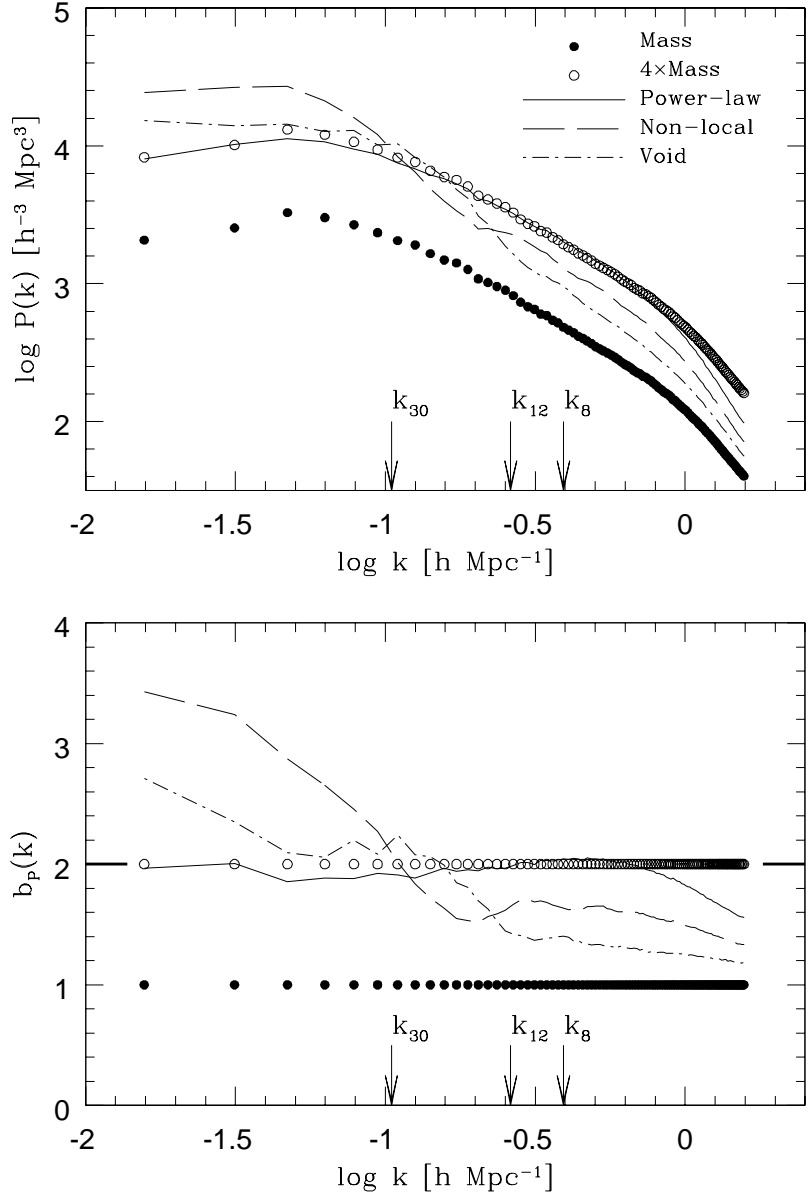}
}
\caption{
Top:  Power spectra of the galaxy distributions under the power-law, 
non-local, and void biasing schemes.
Bottom:  The bias functions $b_{P}(k)$ defined in equation~(\ref{eqn:bkdef}).
\label{fig:nlpk}
}
\end{figure}

Figure~\ref{fig:nlpk} shows the power spectra and the bias functions for the 
non-local and void biasing schemes, again with the local power-law results
shown for comparison.
The arrow marked $k_{30} = 2\pi/(2 \times 30) = 0.1047 h$ Mpc$^{-1}$ 
represents the wavenumber corresponding to $30 \hmpc$, the scale of
influence in the non-local model.
The power spectrum of the non-local model shows strong curvature 
at roughly this scale and dramatically enhanced amplitude on larger 
scales.
The bias function $b_P(k)$ increases with scale
and does not settle to a constant even near the largest scales shown, 
the fundamental frequency of the cubical simulation box.
The void bias model once again exhibits a characteristic feature at the 
scale of the voids, corresponding to the wavenumber 
$k_{v} = 2\pi/(2 \times R_{v}) = 0.2094 h$ Mpc$^{-1}$,
with a large difference in $b_P(k)$ on either side of it.
The non-local and the void bias schemes can both 
enhance the amplitude and change the shape of the underlying mass power 
spectrum, even on very large scales.

\subsection{Higher-Order Moments}	

The correlation function and power spectrum measure second moments of
the galaxy distribution as a function of scale.  We now turn to
some of the simplest measures of higher-order clustering, the
hierarchical amplitudes
\be
S_J(R) \equiv \frac{\langle \delta^J \rangle_c}
		   {\langle \delta^2 \rangle^{J-1}} \; ,
\label{eqn:sj}
\ee
where $\langle \delta^J \rangle_c$ and $\langle \delta^2 \rangle$
are the $J$-th connected moment and the variance, respectively,
of the density contrast field smoothed with a top hat filter
of radius $R$.  The third and fourth connected moments are
$\langle \delta^3 \rangle$ and 
$\langle \delta^4 \rangle - 3\langle \delta^2 \rangle^2$, respectively.
For Gaussian initial conditions, $(J-1)$-th order gravitational
perturbation theory predicts that $S_J(R)$ is independent of the
amplitude of $P(k)$, independent of $R$ for a power-law $P(k)$,
and only weakly dependent on $R$ (i.e., varying much more slowly than
$\langle \delta^J \rangle_c$) for a more general $P(k)$
(\cite{fry84}; \cite{juszkiewicz93}; \cite{bernardeau94}).
A local linear or non-linear transformation of the smoothed field
$\delta({\bf x})$ --- i.e., a form of local bias applied on scale $R$ ---
alters the values of $S_J$, but it does not destroy this 
hierarchical behavior (\cite{fry93}).  The values of $S_J$
can therefore provide a diagnostic for the properties of the
biasing relation.  The hierarchical behavior of moments of the
observed IRAS and optical galaxy distributions
(e.g., \cite{meiksin92}; \cite{bouchet93}; \cite{gaztanaga92}, 1994; 
\cite{kim98}) 
supports the hypothesis of Gaussian initial conditions,
and it has also been used as an argument against non-local bias
(\cite{frieman94}).  

\begin{figure}
\centerline{
\epsfxsize=\hsize
\epsfbox[18 144 592 718]{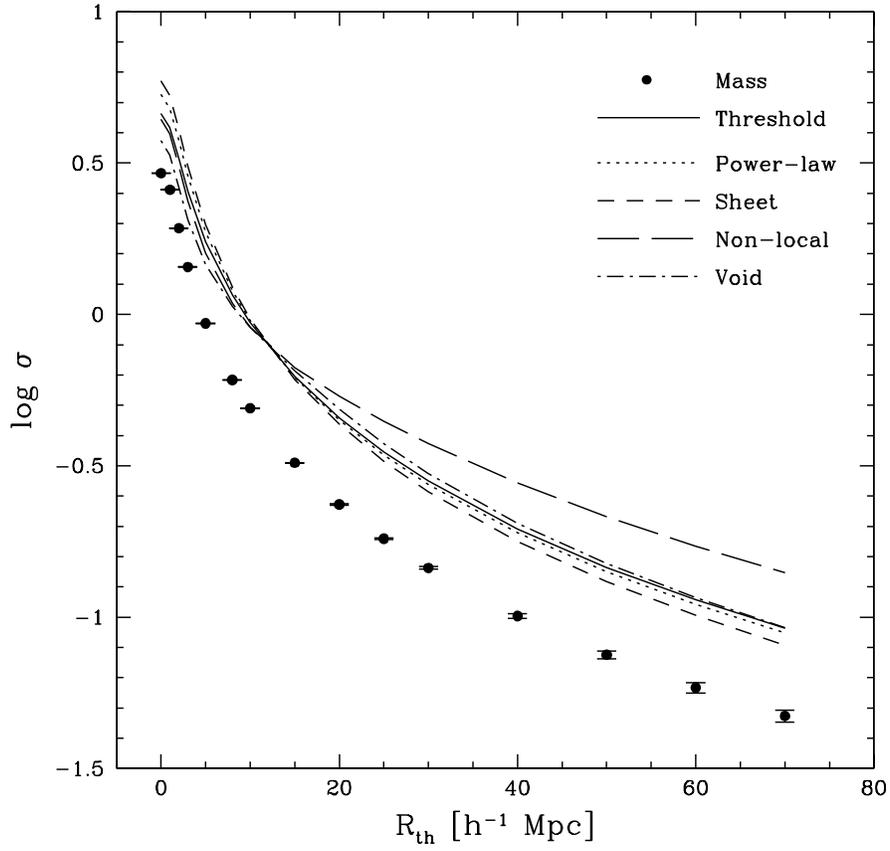}
}
\caption{
The rms fluctuation amplitude $\sigma$ as a function of the radius $R_{\rm th}$
of the top hat filter used to smooth the density fields.
The solid circles show the $\sigma$ of the mass distribution,
while the lines represent the various biased galaxy density fields.
The error bars show the $1\sigma$ deviation computed as the dispersion
in the measurements from four independent simulations and divided by 
$\sqrt{3}$.
\label{fig:rms}
}
\end{figure}

\begin{figure}
\centerline{
\epsfxsize=\hsize
\epsfbox[18 144 592 718]{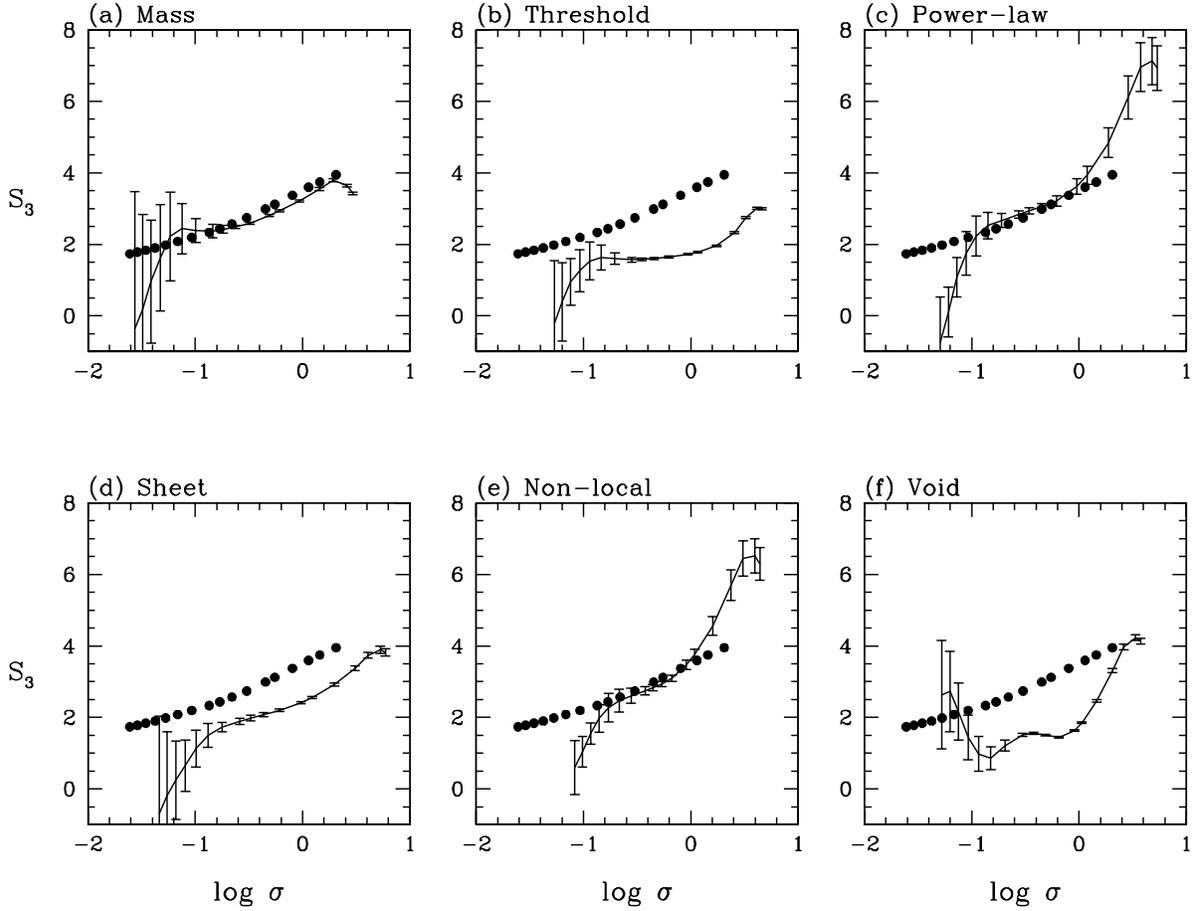}
}
\caption{
The hierarchical amplitude $S_{3}$ as a function of the rms density
fluctuation amplitude $\sigma$ for the mass and biased galaxy distributions.
Solid circles show $S_{3}(\sigma)$ predicted by perturbation theory
for the underlying mass density field.
Error bars show the $1\sigma$ uncertainty computed as the dispersion
in the measurements from four independent simulations divided by 
$\sqrt{3}$.
\label{fig:s3}
}
\end{figure}

\begin{figure}
\centerline{
\epsfxsize=\hsize
\epsfbox[18 144 592 718]{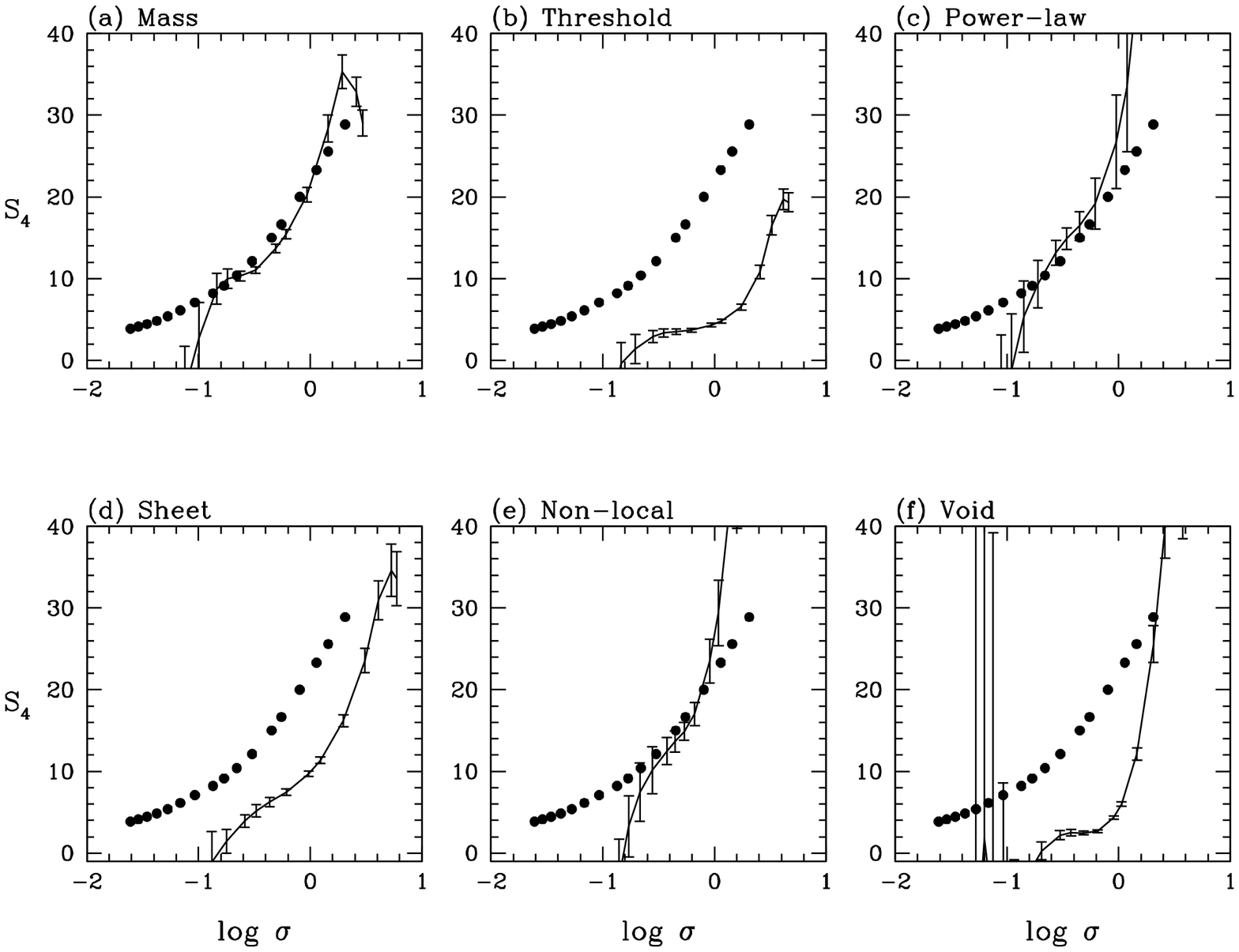}
}
\caption{
The hierarchical amplitude $S_{4}$ as a function of the rms density
fluctuation amplitude $\sigma$ for the mass and biased galaxy distributions.
Solid circles show $S_{4}(\sigma)$ predicted by perturbation theory
for the underlying mass density field.
Error bars show the $1\sigma$ uncertainty computed as the dispersion
in the measurements from four independent simulations divided by 
$\sqrt{3}$.
\label{fig:s4}
}
\end{figure}

Figure~\ref{fig:rms} shows the rms fluctuation 
$\sigma \equiv \langle \delta^2 \rangle^{1/2}$ of
the mass and galaxy density fields as a function of the 
top hat smoothing radius $R_{\rm th}$.
For the sake of clarity, we show error bars only for the mass distribution;
error bars for the galaxy distributions are similar.
We compute the statistical
errors as the dispersion in the measurements from four 
independent simulations, divided by $\sqrt{3}$.
Our normalization condition, $b_\sigma(12)=2$, forces
the rms fluctuations of all models to be equal at $R_{\rm th}=12\hmpc$.
In Figures~\ref{fig:s3} and~\ref{fig:s4}, we plot the hierarchical amplitudes
$S_{3}$ and $S_{4}$ as a function of $\log \sigma$, which can be 
related to a top hat filter radius using Figure~\ref{fig:rms}.
We compute all moments by CIC-binning the mass or galaxy distributions
onto a $200^{3}$ grid and smoothing this density field
with top hat filters of successively larger radii.

Solid lines in Figure~\ref{fig:s3} show $S_3$ for the various bias
models, with error bars estimated from the dispersion of $S_3$
values measured from the four independent simulations, divided
by $\sqrt{3}$.  The filled circles 
are the same in all the panels and show the values of
$S_{3}$ predicted by perturbation theory for the underlying mass 
distribution, computed for our tilted CDM power spectrum
using the equations in \S3.5 of Bernardeau (1994).
The predictions agree very well with the values measured from the
underlying mass distribution (panel a), demonstrating that  the scale 
dependence of $S_{3}$ of the mass distribution can be reproduced quite
accurately using perturbation theory if the matter power spectrum
is known {\it a priori}.
This Figure also shows that the finite size of the simulation box
leads to unreliable estimates of $S_{3}$ for $\log\sigma_{g} \la -0.8$,
corresponding to $ R_{\rm th} \approx 45 \hmpc$.

A linear bias $\delta_g=b\delta$ would reduce the amplitude of $S_3$
by a factor of $b$ on all scales.  Our bias models, both local
and non-local, have a more complicated effect, altering the scale-dependence
of $S_3$.
Density-threshold and sheet bias (panels b and d)
significantly reduce the amplitude of $S_{3}$ on all scales,
and for density-threshold bias $S_3$ becomes almost scale-independent
over the range $0.1 < \sigma < 1$.
The high weight assigned to dense regions boosts the value of $S_3$
on small scales in the power-law bias model (panel c),
though in the range $0.1 < \sigma < 1$ this model tracks the
behavior of the mass distribution remarkably closely.
Values of $S_3(\sigma)$ for the non-local bias model (panel e) are similar
to those for the local power-law bias.
Finally, void bias (panel f) produces a systematically lower $S_3(\sigma)$
and a feature at $\sigma = 0.25$, corresponding to a physical
scale close to the diameter of the voids.

Figure~\ref{fig:s4} shows $S_4(\sigma)$ in the same format as
Figure~\ref{fig:s3}.
The perturbation theory predictions, based on
the equations of Bernardeau (1994), again accurately match
the values measured from the mass distribution over the range 
$0.1 < \sigma < 1$.
Linear bias would decrease $S_{4}$ by a factor of $b^{2}$ on all scales.
The relative behavior of $S_{4}(\sigma)$ 
for the different biasing schemes is similar
to that of $S_{3}(\sigma)$.
Thus, density-threshold bias reduces both the amplitude and the 
scale-dependence of $S_{4}$, while sheet bias primarily reduces its amplitude.
The power-law and the non-local biasing schemes preserve the $S_{4}$ of the 
mass distribution over the range $0.1 < \sigma < 1$ and depart drastically
from it outside this range.
Void bias once again produces a break at $\sigma \approx 0.25$.
Note that although Figures~\ref{fig:s3} and~\ref{fig:s4} demonstrate
the scale-dependence of $S_3$ and $S_4$, all of our biasing models
(even the non-local and void bias)
still preserve the basic hierarchical pattern of moments predicted
by gravitational perturbation theory with Gaussian initial conditions;
for example, the ratios $S_3$ and $S_4$ change by less (usually much less)
than a factor of two between $\log\sigma=-0.5$ and $\log\sigma=0$,
even though the denominators $\sigma^4$ and $\sigma^6$ change by
factors of 100 and 1000, respectively.

Our results provide two cautionary notes for efforts to infer
the bias relation from measurements of hierarchical amplitudes.
First, the agreement between observed amplitudes and those
predicted for the mass distribution has been used to argue
that galaxy formation is unbiased (e.g., \cite{gaztanaga94a}).
However, we find that the power-law bias model, which is not
particularly contrived, happens to yield nearly the same results
as the mass distribution, at least for $S_3$ and $S_4$.
Second, the agreement of $S_3(\sigma)$ and $S_4(\sigma)$
with perturbation theory predictions has been used to argue
against the BCFW93 cooperative galaxy formation bias model
(\cite{frieman94}).  However, the non-local model adopted here
gives $S_3(\sigma)$ and $S_4(\sigma)$ results similar to those
of the local power-law bias model (and the mass distribution),
suggesting that the failure of the BCFW93 model may not extend 
generally to all similar non-local models.

\subsection{Quadratic Bias Parameters}	

Ratios of the moments of galaxy counts can also be interpreted in terms
of a hierarchy of bias parameters describing the relation between
galaxies and mass (\cite{fry93}; \cite{juszkiewicz95}).
Suppose that there is a relation $\delta_g=f(\delta_m)$ between the
galaxy and mass density contrast fields after both are smoothed over
scale $R$.  In the limit of small amplitude fluctuations,
$|\delta_m|\ll 1$, the function $f(\delta_m)$ can be approximated
by a second-order Taylor expansion,
\be
\delta_{g} \equiv f(\delta_{m}) \simeq b_{1}\delta_{m} + 
    \frac{1}{2}b_{2}\delta_{m}^{2} - \frac{1}{2}b_{2}\sigma_{m}^{2},
\label{eqn:taylor}
\ee
where the third term ensures that $\langle \delta_{g} \rangle = 0$.
The ``linear'' bias parameter $b_1$ gives the slope of the galaxy-mass
relation, and in the limit of first-order perturbation theory it is 
equal to other bias parameters such as $b_\sigma$ and $b_\xi$.
The Taylor expansion~\eqref{taylor} can be regarded as the justification
for adopting the linear bias model for some calculations
in the limit of small $\delta_m$.
However, the hierarchical amplitude $S_3$ is only non-zero in
second-order perturbation theory, so to compute the effect of bias
on $S_3$ one must use the full second-order expansion~\eqref{taylor}
even in the small-$\delta_m$ limit.
The result is (\cite{fry93}; \cite{juszkiewicz95})
\be
S_{3g} =  \frac{S_{3m}}{b_{1}} + \frac{3b_{2}}{b_{1}^{2}}.
\label{eqn:s3gdef}
\ee
At the same order, the variance of the galaxy density is
(R. Scoccimarro, private communication)
\be
\sigma_{g}^{2} = b_{1}^{2}\sigma_{m}^{2} + S_{3m}b_{1}b_{2}\sigma_{m}^{4} + 
\frac{1}{2}b_{2}^{2}\sigma_{m}^{4}.
\label{eqn:b12def}
\ee

\begin{figure}
\centerline{
\epsfxsize=\hsize
\epsfbox[18 144 592 718]{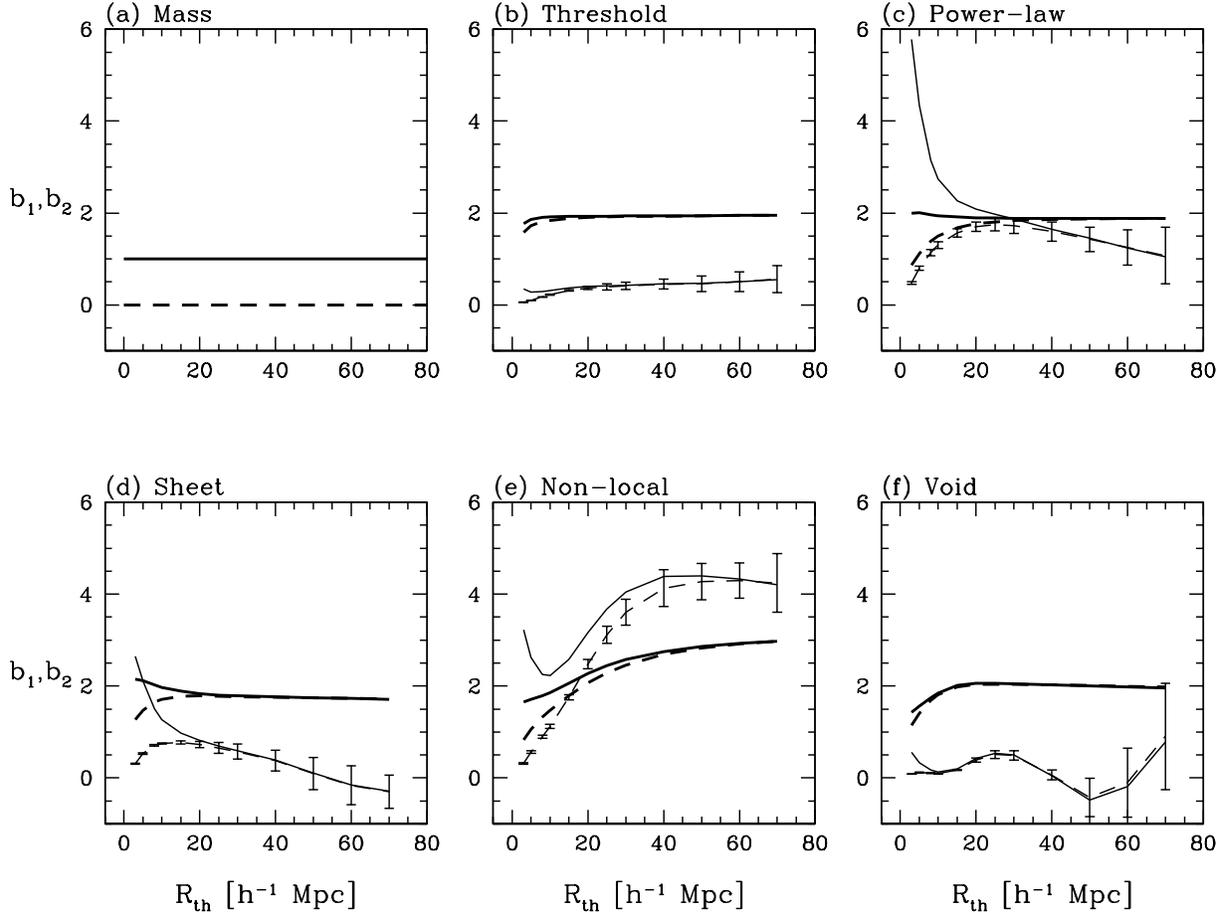}
}
\caption{
The linear and quadratic bias parameters $b_{1}$ and $b_{2}$,
as a function of the top hat smoothing radius $R_{\rm th}$.
Thick and thin solid lines show $b_1 \equiv b_\sigma$
(eq. [\ref{eqn:bdef}]) and $b_2$ computed from equation~\eqref{s3gdef}.
Thick and thin dashed lines show $b_1$ and $b_2$ computed
by simultaneously solving equations~\eqref{s3gdef} and~\eqref{b12def}.
The $1\sigma$ error bars on $b_{2}$ are computed using the dispersion 
in the measurements from four independent simulations, 
divided by $\sqrt{3}$.
\label{fig:b12}
}
\end{figure}

In general, there will be some scatter about the relation 
$\delta_g=f(\delta_m)$, and even if the scatter is small at
one scale it could be large at another scale.
We will return to this issue in \S 3.9 below.
For now, however, we investigate the behavior of the
linear and quadratic bias parameters $b_1$ and $b_2$ 
{\it defined} as the simultaneous solutions to equations~\eqref{s3gdef}
and~\eqref{b12def}.
The thick and the thin dashed lines in 
Figure~\ref{fig:b12} plot these parameters
as a function of $R_{\rm th}$, the radius of the top hat 
filter used to define the smoothed density fields.
The thick and thin solid lines show the results of an alternative
definition in which equation~\eqref{b12def} is replaced by the
first-order relation $b_1 = b_\sigma = \sigma_g/\sigma_m$ 
(eq.~[\ref{eqn:bdef}]).  The two definitions are identical in
the limit $\sigma_m \ll 1$, since the $\sigma_m^4$ terms in
equation~\eqref{b12def} become negligible.
We estimate the errors in $b_{2}$ as the dispersion in the values
derived from the four independent simulations, divided by $\sqrt{3}$.
The errors in $b_{1}$ are tiny over the range of 
scales considered, so we do not show them in the Figure.

The rms fluctuation $\sigma$ is given by an integral of $\xi(r)$.
Since we have already shown that $b_{\xi}(r)$ tends to a constant on 
large scales in local bias models, we expect $b_1$ to become constant
on large scales in local models as well.  This is just the behavior
that we find in Figure~\ref{fig:b12}, though in the sheet bias model
there is scale-dependence out to $R \sim 15\hmpc$.
The void bias model has a more scale-dependent value of $b_1$,
though it still settles to $b_1=2$ for $R \geq 20\hmpc$.
In the non-local model, on the other hand, $b_1$ increases 
monotonically with scale up to the largest scales plotted,
reaching $b_1=3$ at $R_{\rm th}=70\hmpc$.  In all cases the
definition $b_1=b_\sigma$ is less scale-dependent than the
definition from simultaneous solution of equations~\eqref{s3gdef}
and~\eqref{b12def}.

Given that $b_1$ and $b_\xi$ become scale-independent on large scales
in all of our local biasing models, it is tempting to  conjecture that 
the quadratic bias parameter $b_{2}$ also becomes scale-independent
in this regime.  Indeed, one might 
extend this conjecture to scale-independence
of the whole hierarchy of non-linear bias parameters defined by
Fry \& Gazta\~naga (1993), which come from the successively higher order
terms in the Taylor expansion of $\delta_g=f(\delta_m)$.
Unfortunately, the noise in estimates of $S_3$ in our finite simulation
box makes it difficult to test even the $b_2$ conjecture.  The density-threshold
and power-law bias results appear marginally consistent with a constant
$b_2$ for $R_{\rm th}>15\hmpc$, while the sheet bias model appears
marginally inconsistent with constant $b_2$.  
The value of $b_2$ certainly shows more scale-dependence at large 
$R_{\rm th}$ in the non-local and void bias models.  
The value of $b_2$ is scale-dependent in all models except 
density-threshold bias for $R_{\rm th}<15\hmpc$,
which is not too surprising since the Taylor expansion~\eqref{taylor}
and perturbation theory calculation~\eqref{s3gdef} break down
as $\sigma$ approaches one.

\subsection{Pairwise Peculiar Velocities}

Gravitational evolution of the inhomogeneous mass distribution induces
peculiar velocities on all the mass particles.
The pairwise velocity dispersion, which can be estimated from
the anisotropy of the redshift-space correlation function
$\xi(r_p,\pi)$ of galaxy redshift surveys (\cite{dp83}; \cite{bean83}),
provides a diagnostic of $\Omega$ by way of the ``Cosmic Virial Theorem''
(\cite{peebles76}).  It has also been used as a direct test
of cosmological models (e.g., \cite{davis85}).
The first and second moments of the pairwise velocity distribution
enter into the BBGKY equations (\cite{davis77}) and analytic
calculations of redshift-space distortions of $\xi(r_p,\pi)$
(\cite{fisher95} and references therein).
However, even if galaxies have the same velocity field as the
dark matter, these statistical characterizations of galaxy velocities
can be strongly affected by bias because of the pair weighting.
For example, if biased galaxy formation preferentially populates 
dense regions with high velocity dispersions, then the pairwise
dispersion of the galaxies will exceed the pairwise dispersion
of the mass.

\begin{figure}
\centerline{
\epsfxsize=\hsize
\epsfbox[18 144 592 718]{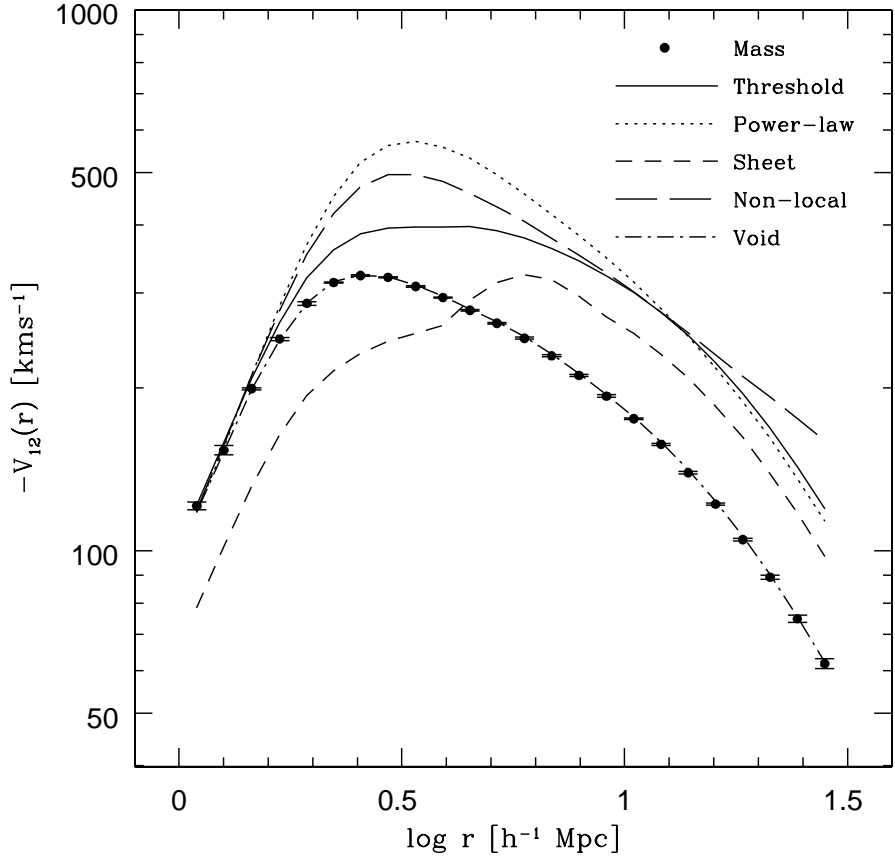}
}
\caption{
The mean pairwise radial peculiar velocity $V_{12}(r)$ as a function 
of pair separation for the underlying mass distribution and the various biased 
galaxy distributions.
Error bars, usually smaller than the points, are plotted
for the mass distribution only; they are similar
in magnitude for all the galaxy distributions.
We plot $V_{12}$ logarithmically so that a constant offset
between galaxy and mass curves corresponds to a constant amplification
of $V_{12}(r)$, as expected for linear bias.
\label{fig:v12}
}
\end{figure}

Figure~\ref{fig:v12}
shows the first moment of the pairwise velocity distribution,
the mean pairwise radial velocity 
\be
V_{12}(r) \equiv \langle ({\bf v}_{1}-{\bf v}_{2}) \cdot {\bf r}_{12} \rangle .
\label{eqn:v12def}
\ee
Here ${\bf v}_{1}$ and ${\bf v}_{2}$ are the velocities of two 
particles separated by the vector ${\bf r}_{12}$,
and the angular bracket denotes an average 
over all particle pairs with separation $r = \vert {\bf r}_{12} \vert$.
Error bars are estimated from the dispersion among the four independent 
simulations, divided by $\sqrt{3}$.
In a linear bias model $\delta_g=b\delta_m$ where the galaxies
follow the same velocity field as the mass,
$V_{12}^{\rm gal}(r)=bV_{12}^{\rm mass}(r)$ in the 
small $\delta_m$ (large $r$) limit (\cite{fisher94}).
In line with this expectation, all three of our local bias
models exhibit a nearly constant amplification of $V_{12}(r)$
for $r > 10\hmpc$, by a factor that is close to the model's
value of $b_\xi(r)$ from Figure~\ref{fig:xi}.  
At smaller scales, however, the shape of $V_{12}(r)$ depends 
strongly on the biasing scheme, in particular on the degree
to which it weights the densest regions.  Thus, the power-law
bias model has the highest $V_{12}(r)$ and the sheet model,
which avoids the isotropic clusters, has a $V_{12}(r)$ that
falls below the mass $V_{12}(r)$ at small separations.
The non-local model roughly follows the power-law model
on small scales, and it does not settle to a constant amplification
of $V_{12}(r)$ at large scales.  The ``bias factor'' defined by
$b_v(r)=V_{12}^{\rm gal}(r)/V_{12}^{\rm mass}(r)$ for this model
has a scale-dependence at large $r$ that is reminiscent of
(but not as strong as) that of $b_\xi(r)$ (Figure~\ref{fig:nlxi}).
The void bias model is perhaps the oddest counterexample to the
linear bias expectation: since galaxies are eliminated from randomly
placed voids with no regard to their density or velocity, void bias
does not alter $V_{12}(r)$ at all.  This result emphasizes a
difference between the void bias model and our other models,
namely that it enhances $\xi(r)$ and $P(k)$ by imprinting an
additional, independent clustering pattern on the galaxy distribution
rather than by preferentially selecting galaxies in clustered regions.

\begin{figure}
\centerline{
\epsfxsize=\hsize
\epsfbox[18 144 592 718]{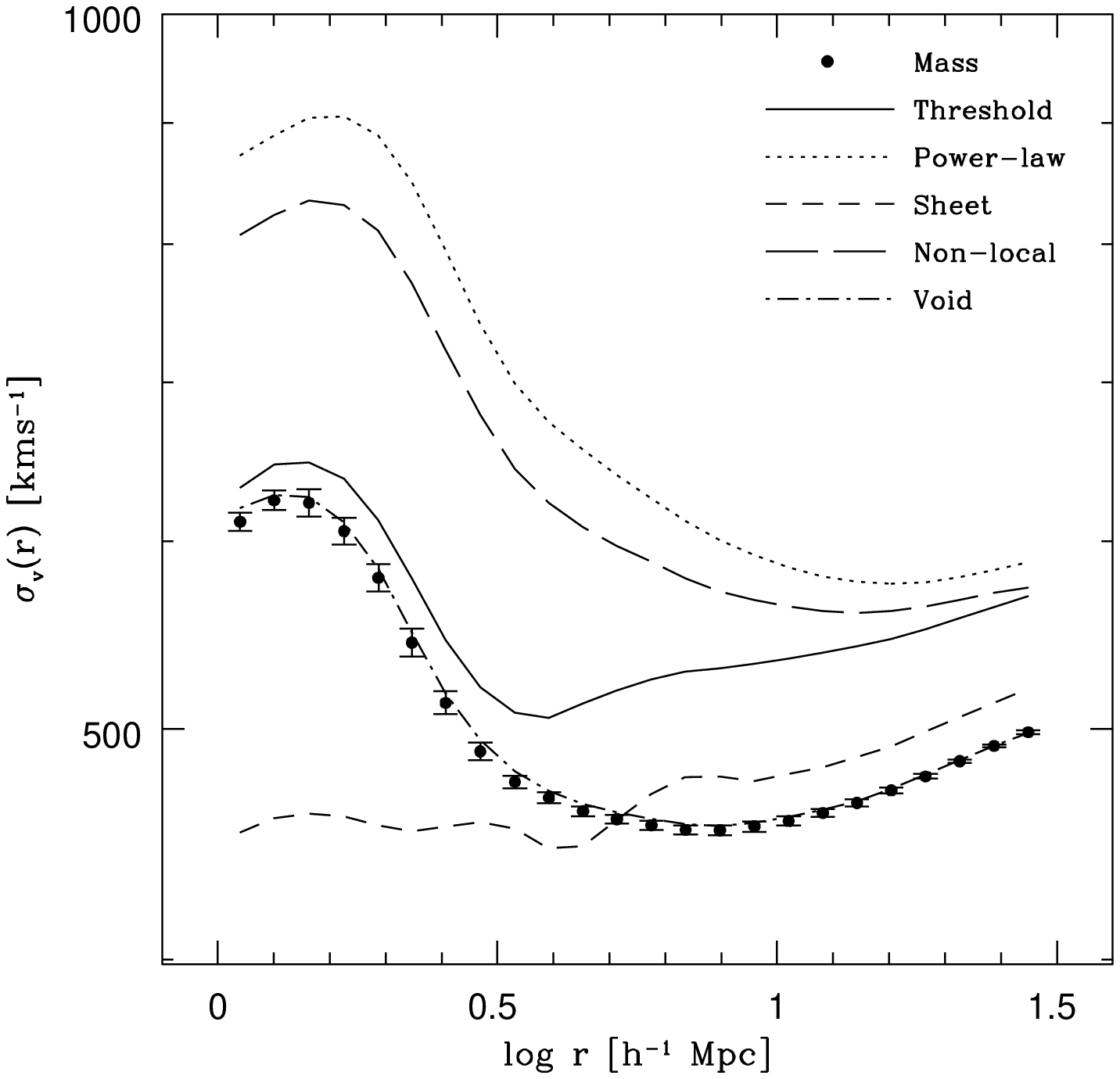}
}
\caption{
The mean pairwise radial peculiar velocity dispersion as a function of the 
pair separation for the mass and the various biased galaxy distributions.
Error bars are shown only for the mass distribution and are similar
in magnitude for all the galaxy distributions.
\label{fig:sigmav}
}
\end{figure}

Figure~\ref{fig:sigmav} shows the second moment of the pairwise
velocity distribution, the pairwise radial velocity dispersion 
\be
\sigma_{v}^{2}(r) \equiv 
\langle \vert ({\bf v}_{1}-{\bf v}_{2}) \cdot {\bf r}_{12} \vert ^{2} \rangle 
- V_{12}^2(r) ,
\label{eqn:sigmardef}
\ee
where the average is again over pairs of separation $r=|{\bf r}_{12}|$.
Measurements of this quantity from the anisotropy of $\xi(r_p,\pi)$
are sensitive to the presence or absence of rich clusters in the
redshift sample (\cite{mo93}; \cite{zurek94}; \cite{somerville97}).
Figure~\ref{fig:sigmav} illustrates another shortcoming of
the pairwise velocity dispersion as a cosmological diagnostic:
it is sensitive to the details of the biasing scheme, so it
cannot be predicted without a fully specified model of biasing.
The two models with the steepest small scale correlation functions,
sheet bias and power-law bias, have, respectively, the lowest and
highest pairwise velocity dispersions at small scales because
of the relative weights they assign to rich clusters.
The sheet bias model has $\sigma_v^2(r)$ well below the mass
$\sigma_v^2(r)$ for $r<4\hmpc$.  At large $r$, all bias models
except for the void model produce an amplification of $\sigma_v^2(r)$,
but not by a factor that is simply related to the large scale bias.
Because many of the pairs contributing high values of the pairwise
velocity at large $r$ have one member in a high-dispersion cluster,
the pairwise dispersion statistic is heavily influenced by
cluster weighting even at large scales.

\begin{figure}
\centerline{
\epsfxsize=\hsize
\epsfbox[18 144 592 718]{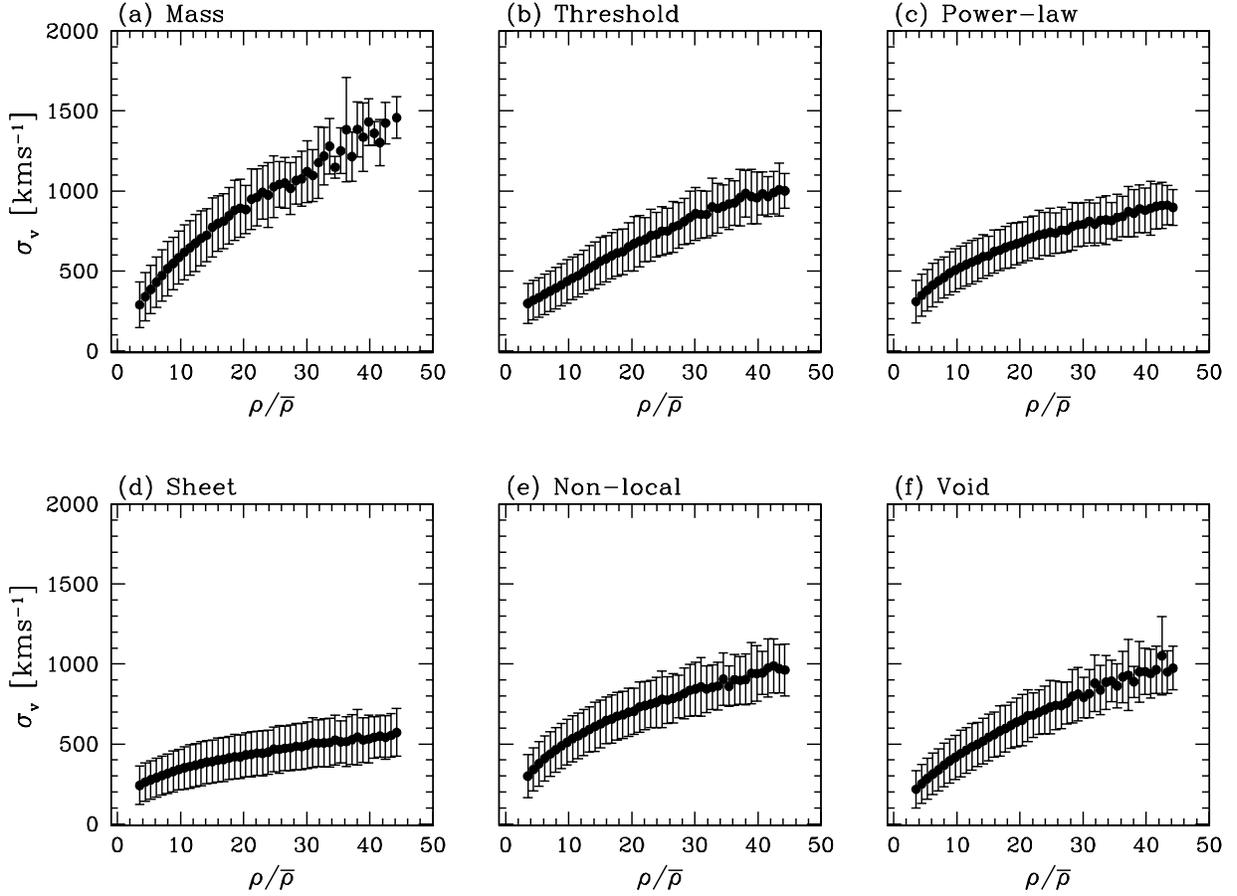}
}
\caption{
Velocity dispersion $\sigma_v$ as a function of the local mass (a) or
galaxy (b-f) overdensity $\overden$.
Both quantities are computed in spheres of radius $3 h^{-1}$Mpc
using four independent simulations.
Error bars show the intrinsic dispersion in $\sigma_{v}$ at fixed
$\overden$, not the (small) statistical uncertainty in our
estimate of the mean $\sigma_v(\overden)$.
\label{fig:sigmavrho}
}
\end{figure}

Recognizing the sensitivity of the pairwise velocity dispersion 
to the presence of clusters in redshift surveys and to the nature of bias,
Kepner, Summers, \& Strauss (1997) and Strauss, Ostriker, \& Cen (1998)
have proposed alternative statistics that measure
the velocity dispersion as a
function of the local number density of galaxies instead
of averaging over all pairs.
(Davis, Miller, \& White [1997] and Landy, Szalay, \& Broadhurst [1998]
have explored other approaches that measure a globally averaged
quantity but do not weight by galaxy pairs.)
In particular, Kepner et al. (1997) suggested that the slope of the 
relation between the velocity dispersion and the galaxy 
number density could provide a diagnostic for $\Omega$.
Figure~\ref{fig:sigmavrho} shows $\sigma_v(\overden)$,
the velocity dispersion as a function of the local galaxy overdensity,
with both quantities computed in spheres of radius $3 \hmpc$.
We also compute the intrinsic dispersion in $\sigma_v$ at fixed
$\overden$, which is shown by the error bars.
This measure of $\sigma_v(\overden)$ is similar but not identical
to the measures proposed by Kepner et al.\ (1997) and
Strauss et al.\ (1998).
Figure~\ref{fig:sigmavrho} shows that the slope of $\sigma_v$ vs.
$\overden$ is systematically lower for biased galaxy distributions
than for the underlying mass distribution, as anticipated by
Kepner et al.\ (1997) and Strauss et al.\ (1998).
More significantly, the slope is not simply related to the rms bias
on the $3\hmpc$ measurement scale (see Figure~\ref{fig:rms})
but depends on the details of the adopted biasing scheme.
Measuring $\sigma_v$ as a function of density does soften the
extreme sensitivity of the pairwise dispersion to the details
of the bias model, but probably not enough to make this statistic
a good one with which to measure $\Omega$.
Measurements of $\sigma_v(\overden)$ can provide a test of cosmological
models, but only if the relation between galaxies and mass
is reliably specified.

\subsection{Galaxy and Mass Density Fields}

Any physical biasing of the galaxy distribution can lead to a 
non-trivial relation between the galaxy and the mass density fields.
We have introduced several different definitions of bias factors,
each related to a specific statistical measure of clustering.
A more general description of the effect of bias is the conditional
probability $P(\delta_g|\delta_m)$ of finding galaxy density contrast
$\delta_g$ where the mass density contrast is $\delta_m$ 
(\cite{dekel98}).  This formulation implicitly assumes a smoothing
scale on with the fields $\delta_g$ and $\delta_m$ are defined,
and in general both the mean relation and the scatter about it will depend
on the smoothing scale.

\begin{figure}
\centerline{
\epsfxsize=\hsize
\epsfbox[18 144 592 718]{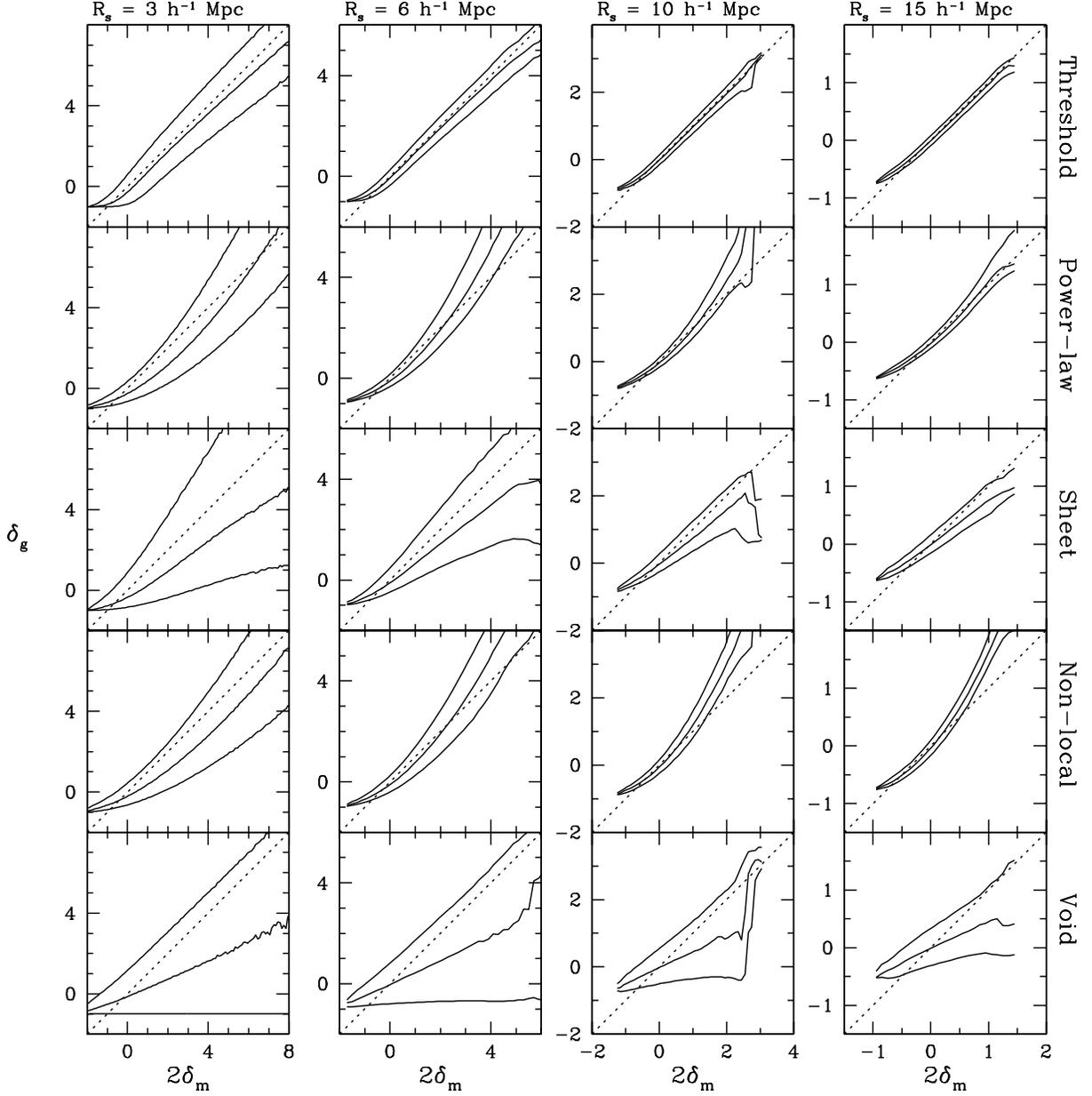}
}
\caption{
The galaxy overdensity $\delta_{g}$ as a function of the 
mass overdensity $\delta_{m}$ for the different biasing schemes
(labeled at right).
In all panels the central line shows the median relation and the
upper and lower lines represent the 90th and the 10th percentile of
the distribution.
From left to right, the different columns show the relation 
when the fields are smoothed with Gaussian filters of radii 
$R_{s} = 3, 6, 10$ and $15 \hmpc$, respectively.
\label{fig:rhorho}
}
\end{figure}

Figure~\ref{fig:rhorho} shows the median and the 
10th and 90th percentile limits
of the distribution of smoothed galaxy density contrasts $\delta_g$
at given values of the smoothed mass density contrast $\delta_m$.
The dotted line that runs along the diagonal in each panel shows the
relation $\delta_g=2\delta_m$ expected for linear bias with a
constant bias factor $b=2$.
Different rows correspond to different biasing schemes,
while the four columns from left to right show results for
Gaussian smoothing radii $R_{s} = 3,6,10$ and $15 \hmpc$, respectively.
We compute the distribution and the different percentile values using the
density fields from the four independent simulations.

The effect of the density-threshold bias is remarkably close to
linear bias at all smoothing scales and density contrasts,
breaking down only where it must at $\delta_g \approx -1$.
This result is consistent with the very small values of $b_2$
found for this model (Figure~\ref{fig:b12}).
The median $\delta_g(\delta_m)$ is markedly more non-linear
for the power-law model, approaching the dotted line only
for a large smoothing length $R_s=15\hmpc$.
The median relation for the sheet model is fairly linear, 
though always shallower than $\delta_g=2\delta_m$.
The non-local model has a noticeably curved median relation 
even at $R_s=15\hmpc$, consistent with the rise in $b_2$ for
this model at large scales (Figure~\ref{fig:b12}).
For the void model, the scatter in $\delta_g$ at fixed $\delta_m$
is so large that the median relation alone provides little information.

More generally, the scatter between $\delta_g$ and $\delta_m$ 
reveals the influence of factors other than the local mass
density in shaping the galaxy distribution.  As one might expect,
the scatter about the median relation is small for the
bias models based on local mass density (density-threshold and
power-law bias), and it decreases rapidly with increasing smoothing.
Some scatter is inevitable because of shot noise in the galaxy distribution,
and because the density smoothed with a Gaussian filter of radius $R_s$
is not perfectly correlated with the density smoothed with a top hat
of radius $R_{\rm th}=4\hmpc$ (as used in the biasing prescriptions).
The influence of large scale density contrast in the non-local
model does not significantly increase the scatter in $\delta_g$ vs.
$\delta_m$ over that in the power-law model.
In the sheet bias model, on the other hand, galaxy selection is 
based on the geometry of the local mass distribution rather than
the mass density.  These quantities are correlated, but the
scatter between $\sigma_g$ and $\sigma_m$ is much larger than 
in the density-based models, though it still becomes fairly small for
$R_s \geq 10\hmpc$.
In the void bias model, the probability that a mass particle is included as 
a galaxy is entirely independent of the local mass density, since the 
voids are thrown at random into the mass distribution.
The void bias model therefore has large scatter between $\delta_g$
and $\delta_m$, persisting to very large scales.
At small $R_s$, the 10th percentile at all $\delta_m$ corresponds
to regions that are empty of galaxies, while the 90th percentile
corresponds to regions in which the smoothing volume does not overlap
any of the voids.

In practice, the mass density contrast $\delta_m$ used as the 
independent variable in Figure~\ref{fig:rhorho} is not directly 
observable.  However, $\delta_m$ can be inferred from the
divergence of the peculiar velocity field using the linear
perturbation theory relation 
$\delta_m = -\Omega^{-0.6} \nabla\cdot {\bf v}/H_0$
or weakly non-linear relations 
(\cite{nusser91}; \cite{gramann93}; \cite{chodorowski97}; 
\cite{susperregi97}).
The quantity $\nabla \cdot {\bf v}$ can be inferred from 
observations of the radial peculiar velocity field 
by the POTENT method, under the assumption that the 
peculiar velocity field remains 
irrotational during gravitational evolution 
(\cite{bertschinger89}; \cite{dekel93}; \cite{sigad98}).
In the case of linear bias, the slope of the relation between 
$\delta_g$ and $\nabla\cdot {\bf v}$ provides
an estimate of the quantity $\beta = \Omega^{0.6}/b$.

\begin{figure}
\centerline{
\epsfxsize=\hsize
\epsfbox[18 144 592 718]{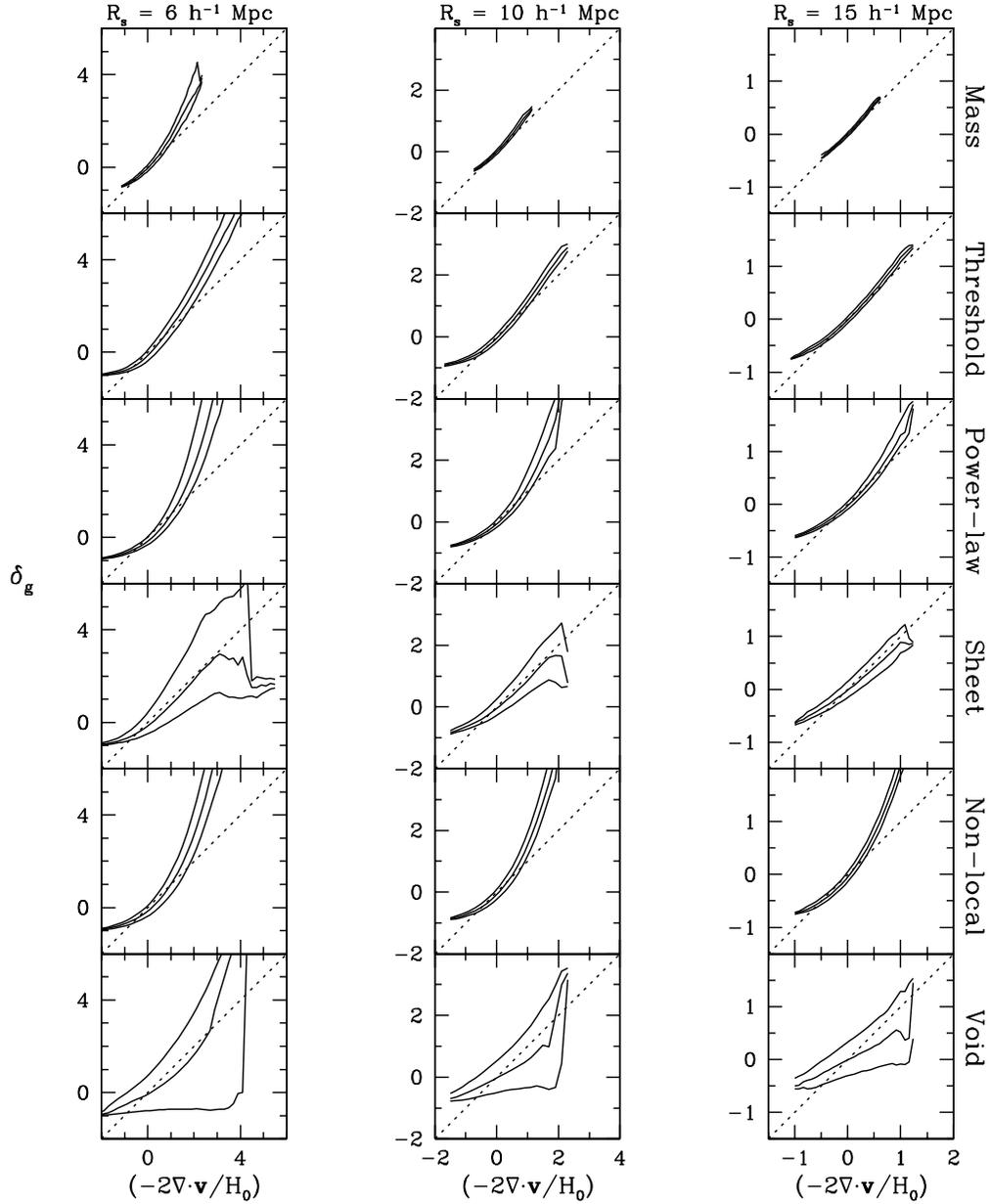}
}
\caption{
The galaxy overdensity $\delta_{g}$ as a function of the negative 
divergence of the velocity field (divided by the Hubble constant),
for the mass distribution and the various biased galaxy distributions
(labeled at right).
In all panels the central line shows the median relation and the
upper and lower lines represent the 90th and the 10th percentile of
the distribution.
From left to right, the different columns show the relation 
when the density and velocity fields are smoothed with 
Gaussian filters of radii 
$R_{s} = 6, 10$ and $15 \hmpc$, respectively.
The first row of panels shows the relation between $\delta_m$
and $-\nabla \cdot {\bf v}/H_0$, while subsequent rows show
the relation between $\delta_g$ and $-2\nabla\cdot {\bf v}/H_0$.
\label{fig:rhodivv}
}
\end{figure}

Figure~\ref{fig:rhodivv} shows the relation between $\delta_g$
and $-\nabla \cdot {\bf v}$ for the unbiased mass distribution
(top row) and the various biased galaxy distributions,
in the same format as Figure~\ref{fig:rhorho}.
The three columns from left to right show the relation when the fields 
are smoothed with Gaussian filters of radius $R_{s} = 6,$ 10, and $15 \hmpc$, 
respectively.
We use the method of Babul et al.\ (1994) to create
a volume-weighted, smoothed velocity field from the discrete galaxy
peculiar velocities.  Specifically, we first
form the momentum field by CIC-binning the momentum of every galaxy
onto a $200^{3}$ grid.
We smooth this momentum field with a Gaussian filter of radius
$R_{1} = R_{s}/2$ and divide it by a similarly smoothed density field to 
form a mass-weighted smoothed velocity field.
We then smooth this velocity field with another Gaussian filter of 
radius $R_{2} = (R_{s}^{2} - R_{1}^{2})^{1/2}$, so that the 
effective smoothing radius is $R_s$.
Because the second smoothing dominates over the first, the
final velocity field is volume-weighted rather than mass-weighted.
We do not show the $\delta_v$ vs. $-\nabla\cdot {\bf v}$ relation
at $R_{s} = 3 \hmpc$ because the discrete galaxy distribution
yields an excessively noisy velocity field in low density regions
for this smoothing length.

In the first row of Figure 17, systematic departures of the solid curves
from the linear theory relation 
$\delta_{m} = - \nabla \cdot {\bf v}/H_{0}$ are caused by
non-linear gravitational evolution.
This deviation decreases drastically with increased smoothing,
showing that linear theory is an increasingly better approximation
of the gravitational evolution on larger scales.
The panels in the remaining rows of Figure 17 show the distribution of the 
galaxy density fluctuations $\delta_{g}$ plotted against 
$-2 \nabla \cdot {\bf v}/H_{0} $.
For $R_s=15\hmpc$ these plots look similar to the corresponding
panels of Figure~\ref{fig:rhorho}.  For smaller smoothing
lengths the relation between $\delta_g$ and $-\divv$ is more non-linear
than the relation between $\delta_g$ and $\delta_m$ because
of the additional non-linearity of the $\delta_g-\delta_m$ relation.
Comparisons between galaxy density fields and mass density fields
estimated from POTENT usually account for the non-linear relation
between $\delta_m$ and $-\divv$, but the non-linearity and scatter
of the bias relation itself are potential sources of additional
systematic error in POTENT estimates of $\beta$, as noted by DL98.
The importance of these effects depends in detail on the form of bias ---
e.g., density-threshold bias produces a nearly linear relation with
small scatter, power-law bias produces a non-linear relation with
small scatter, and sheet bias produces a nearly linear relation
with large scatter.

\section{Discussion}

We have examined the influence of the morphology-density relation
and biased galaxy formation on many of the statistical measures
that are commonly used to characterize galaxy clustering and galaxy
peculiar velocity fields.  We have focused most of our attention on local
biasing models, in which the efficiency of galaxy formation is governed
by the density (density-threshold or power-law bias), geometry
(sheet or filament bias), or ``pressure'' ($\rho\sigma_v^2$) within
a sphere of radius $4\hmpc$.  We have contrasted the behavior of these
local models with that of models in which the galaxy density is
coherently modulated over large scales (non-local bias, as proposed
by BCFW93) or suppressed in randomly distributed voids (void bias,
as proposed by BW91).  In this Section, we summarize our results,
then discuss them in light of other recent work and anticipated
observational developments.

If the morphological segregation of galaxies is governed by the 
local morphology-density relation proposed by Postman \& Geller (1984),
then on small scales the correlation function of early-type galaxies
should be both steeper and stronger than that of late-type galaxies,
in accord with present observations.  On scales larger than the
galaxy correlation length, early-type galaxies should remain more strongly
clustered than late-type galaxies, but the correlation functions (and
power spectra) should have the same shape.  This prediction of the
local morphology-density model is consistent with current data and
can be tested at high precision by the 2dF and SDSS redshift surveys.

In all of the examples that we have investigated, local bias
produces a scale-independent amplification of the two-point correlation
function and power spectrum on scales in the linear regime.
On these scales, therefore, local biasing cannot resolve a discrepancy
between the shape of the mass power spectrum predicted by a cosmological
model and the shape of the observed galaxy power spectrum.
Non-local biasing can resolve such a discrepancy, as originally shown
by BW91 and BCFW93 in the context of ``standard'' CDM and the APM
galaxy survey, but achieving this alteration of the power spectrum shape
requires a bias mechanism that directly modulates galaxy formation 
in a coherent way over large scales.  In the non-linear regime,
by contrast, the bias functions $b_\xi(r)$, $b_\sigma(r)$, and
$b_P(k)$ are scale-dependent even in local models, and their shapes
depend on the details of the biasing scheme.

Local bias models roughly preserve the hierarchical relations between
moments of the galaxy count distribution, in that the ratios
$S_3 \equiv \langle \delta_g^3 \rangle_c/\langle \delta_g^2 \rangle^2$ and
$S_4 \equiv \langle \delta_g^4 \rangle_c/\langle \delta_g^2 \rangle^3$ 
are only weakly dependent on smoothing scale (and hence on
$\sigma \equiv \langle \delta_g^2 \rangle^{1/2}$).  
However, local bias can change both the amplitude and shape
of the functions $S_3(\sigma)$ and $S_4(\sigma)$.
Our power-law bias model gives $S_3(\sigma)$ and $S_4(\sigma)$
close to those of the underlying mass distribution, and our non-local
bias model in turn gives $S_3(\sigma)$ and $S_4(\sigma)$ close to 
those of the local power-law model.  These results show that agreement
between measured hierarchical amplitudes and perturbation theory
predictions for the mass distribution does not necessarily imply
that galaxy formation is unbiased or even that it is locally biased,
although such agreement can rule out specific local and non-local models
(\cite{frieman94}).

If we characterize the effect of bias on the variance and skewness
of galaxy counts by linear and quadratic bias parameters $b_1$ and
$b_2$, then $b_1$ is independent of scale on large scales in all
of our local models.  Our results are marginally but not strongly
inconsistent with the conjecture that $b_2$ becomes scale-independent
on large scales in local bias models.  The scale-dependence of $b_1(r)$
and $b_2(r)$ is much stronger in our non-local model than in any of
our local models.  On large scales, void bias produces a scale-independent
$b_1(r)$ but a scale-dependent $b_2(r)$.

Estimates of moments of the velocity distribution from galaxy
redshift surveys, which are weighted by galaxy pairs, are strongly
affected by biasing because of correlations between the velocity
distribution and the parameters that determine the galaxy formation
efficiency.  In particular, the pairwise velocity dispersion $\sigma_v(r)$
is sensitive to the details of the bias model at all separations.
The two bias models with the steepest small scale correlation function,
power-law bias and sheet bias, have values of $\sigma_v$ that differ
by a factor of two at $r \sim 2\hmpc$, because the first enhances
galaxy formation in dense, isotropic clusters while the second does not.
The dependence of the mean pairwise velocity $V_{12}(r)$ on bias is
similarly complex on small scales, but the behavior simplifies for
local bias models at large $r$, where the galaxy $V_{12}$ is amplified
over the mass $V_{12}$ by a factor $b_v$ that is close to the correlation
function bias $b_\xi$.  These biases of pairwise moments arise even though 
the galaxies in our models are just a subset of the dark matter particles
and therefore have the same local velocity distribution.
Void bias presents an odd case in which the galaxy distribution
is biased but $\sigma_v(r)$ and $V_{12}(r)$ are not.
The relation between local velocity dispersion and local overdensity
(Kepner et al.\ 1997; Strauss et al.\ 1998) is also sensitive
to bias, complicating the use of this statistic as a diagnostic for $\Omega$.

The median trend and scatter of the relation between $\delta_g$ and
$\delta_m$ or $\delta_g$ and $-\divv$ depends on the biasing scheme
and on the smoothing scale used to define the density and velocity fields.
The density-threshold bias prescription produces a relation that is remarkably
close to linear bias, $\delta_g=b\delta_m$.  Power-law bias produces
a curved $\delta_g-\delta_m$ relation, and, as noted by DL98, 
the non-linearity of this
relation is a possible source of systematic error in efforts to
measure $\beta \equiv \Omega^{0.6}/b$ via the POTENT method
(\cite{dekel93}; \cite{sigad98})
or via redshift space distortions (\cite{hamilton98} and references
therein).  The relation between $\delta_g$ and 
$\delta_m$ is fairly tight for the bias schemes that are based
on local density, but it exhibits much more scatter for the sheet
bias scheme, as one might expect.  The void bias model predicts
very large scatter between $\delta_g$ and $\delta_m$ or $-\divv$,
even for smoothing lengths as large as $15\hmpc$.  The observed correlation
between $\delta_g$ and $-\divv$ (\cite{sigad98}) is probably sufficient
to rule out such a model, as first argued by Babul et al.\ (1994).
However, our less extreme non-local bias model predicts a relation
that is nearly as tight as that of the local power-law bias model,
so the existence of a tight relation between $\delta_g$ and $-\divv$
is not sufficient to rule out non-locally biased galaxy formation.

Our results for the large scale behavior of $\xi(r)$ and $P(k)$
strengthen the conclusions of earlier analytic arguments 
(\cite{coles93}; \cite{fry93}; \cite{gaztanaga98}; \cite{scherrer98})
and numerical investigations (\cite{weinberg95}; MPH98; \cite{cole98}).
Although we can only examine a finite number of specific biasing
prescriptions, these examples show that scale-independent amplification
of $\xi(r)$ and $P(k)$ occurs even in models like sheet, filament,
and pressure bias, where the galaxy formation efficiency is not
governed strictly by the local mass density.  We also find that the
asymptotic regime of nearly constant $b_\xi$, $b_\sigma$, and
$b_P$ is effectively reached on mildly non-linear scales,
soon after $\xi(r)$ drops below one.

Our conclusions may seem mildly at odds with those of Blanton et al.\
(1998, hereafter BCOS98), who find scale-dependent bias of the ``galaxy''
population in a hydrodynamic simulation of the $\Lambda+$CDM model 
(by \cite{co98})
The difference is largely a matter of emphasis: BCOS98 find substantial
scale-dependence of $b_\sigma(r)$ in the non-linear regime,
but they find only a 12\% drop in $b_\sigma(r)$ from $r=8\hmpc$ 
to $r=30\hmpc$, which is the same drop we find for our sheet bias
model.  Our power-law and sheet bias models also show significant
scale-dependence of $b_\sigma(r)$ in the non-linear regime  ($ r < 8\hmpc$),
though not as strong as that found by BCOS98.
BCOS98 demonstrate that the scale-dependence of bias in their
simulation arises mainly from the correlation between galaxy formation
efficiency and the local gas temperature $T$ or 
dark matter velocity dispersion $\sigma_v^2$.  They further
argue that this correlation leads to scale-dependence because of
the connection between $T$ (or $\sigma_v^2$) and the gravitational
potential, which has a much redder power spectrum than the density
field itself.  In the terminology of this paper, the BCOS98 argument
could be rephrased as a claim that temperature is a local variable
whose influence is more ``effectively non-local'' than that of
other local variables.  The fact that we obtain virtually identical
results for density-threshold and pressure bias implies that enhanced
scale-dependence is not an {\it automatic} consequence of incorporating
temperature or $\sigma_v^2$ into the biasing prescription.
In order to address this issue more directly, we also examined a 
model in which we biased the galaxy distribution using a threshold
in $\sigma_v^2$ alone, a pure ``temperature'' bias.  We again
obtained results nearly identical to those of the density-threshold
model, with no enhanced scale-dependence of the bias.
The difference of our result from that of BCOS98's similar numerical
experiment presumably reflects our use of a $4\hmpc$ rather than
a $1\hmpc$ sphere to define $\sigma_v^2$.  Averaged over this 
larger scale, velocity dispersion does not behave any more ``non-locally''
than density.  This result does not mean that galaxy bias in a realistic
model might not be as scale-dependent as BCOS98 find, only that the influence
of temperature or velocity dispersion on the scale-dependence of
bias depends in detail on the scale over which it is defined and the
way that it is incorporated into the bias prescription.

Our numerical study complements the general analytic examination of
stochastic, non-linear biasing by DL98.  The use of N-body simulations
allows us to investigate the effects of a wide range of biasing
prescriptions on measures of galaxy clustering in the linear, mildly
non-linear, and strongly non-linear regimes.  For the most part,
we have addressed different issues from DL98, but we concur on the
general point that bias is a multi-faceted phenomenon, and that only in
specific cases and limits can it be described by a single parameter.
On small scales, the relation between $\delta_g$ and $\delta_m$ is
generically non-linear and scale-dependent, and it may have substantial
scatter.  At any given scale one can define many different ``bias factors'' 
--- $b_\xi$, $b_\sigma$, $b_P$, $b_1$, $b_2$, $b_v$, etc. ---
and the relation among them depends on the details of the biasing
scheme, or, ultimately, on the physics of galaxy formation.

Despite this complexity, our results show that the very general assumption
of local biasing leads to some important simplifications on large
scales.  Most significantly, the scale-independence of $b_\xi$ and
$b_P$ in the linear regime means that large galaxy redshift surveys
like 2dF and the SDSS should reveal the true shape of the dark matter
power spectrum on large scales, if galaxy formation is governed by
local physics.  Indeed, this result makes the local biasing hypothesis
testable, since any non-local physics that significantly modulated
galaxy formation would almost certainly have a different impact on
galaxies of widely differing luminosity, stellar population age,
morphology, and surface mass density.  Existing data clearly show
that different types of galaxies have different clustering amplitudes,
but if each galaxy's properties are determined by the history of its
local environment then the 2dF and SDSS redshift surveys should show
that all galaxies have the same $P(k)$ shape on large scales.

The uncertainties of bias have been a source of frustration in efforts
to test cosmological models against observations of galaxy clustering.
In recent years, observational and theoretical breakthroughs have
opened a number of alternative routes to measuring cosmological parameters
and the mass power spectrum, including microwave background anisotropies,
the Ly$\alpha$ forest, the supernova Hubble diagram, weak gravitational 
lensing, and the mass function and evolution of the galaxy cluster population.
These approaches are insensitive or weakly sensitive to biased galaxy
formation, though each one has its own set of assumptions and limitations.
Recent years have also seen great improvements in the predictive power
of theories of galaxy formation, thanks to advances in numerical simulations
and semi-analytic modeling techniques that combine gravitational
clustering with the more complicated
physical processes of gas cooling, star formation, supernova feedback,
metal enrichment, and morphological transformation via mergers and
interactions.  While the sensitivity of galaxy clustering statistics
to the details of galaxy biasing is an obstacle to testing cosmological
models, it becomes an asset when the goal is testing the theory of
galaxy formation itself, especially if the underlying cosmological model
is tightly constrained by independent observations.  We have already
argued that $P(k)$ and $\xi(r)$ can test the broad hypothesis of
local galaxy formation, and at a greater level of detail we might,
for example, come to view the pairwise velocity dispersion not as a 
tool for measuring $\Omega$ but as a diagnostic for the importance
of mergers in dense environments.  The giant redshift surveys currently
underway will provide superb data sets for such studies, allowing
precise clustering measurements for finely divided subsets
of the galaxy population over a wide range of scales.
Other kinds of data may play an equal or more important role in
determining the material contents of the universe and 
the origin of cosmic inhomogeneity, but measurements 
of large scale structure at high
and low redshift will guide our understanding of the physics that 
transformed primordial dark matter fluctuations
into the universe of galaxies.

This work was supported by NSF Grant AST-9616822 and
NASA Astrophysical Theory Grant NAG5-3111.
We thank Roman Scoccimarro for helpful discussions on
galaxy moments and their relation to non-linear bias factors.

\end{document}